\documentclass[12pt,draftcls,onecolumn] {IEEEtran}
\usepackage{array}
\usepackage{multirow}
\usepackage{cite}
\usepackage{amsmath}
\usepackage{booktabs}
\usepackage{color}
\usepackage{graphicx}
\usepackage{mathrsfs}
\usepackage{tikz}
\usepackage{calc}
\usepackage{amssymb}
\usepackage{threeparttable}
\newcommand{\xmark}

\ifCLASSINFOpdf
\else
\fi
\begin{document}
\renewcommand\arraystretch{1}
\bibliographystyle{ieeetr}

\title{\color{black}{Joint Training of the Superimposed Direct and Reflected Links in Reconfigurable Intelligent Surface Assisted Multiuser Communications}}

\author{Jiancheng~An, Chao~Xu, \emph{Senior Member, IEEE}, Li~Wang, \emph{Senior Member, IEEE}, Yusha~Liu, Lu~Gan, and~Lajos~Hanzo, \emph{Fellow, IEEE}
	\thanks{This work was supported in part by the China Scholarship Council. L. Hanzo would like to gratefully acknowledge the financial support of the Engineering and Physical Sciences Research Council projects EP/N004558/1, EP/P034284/1, EP/P034284/1, EP/P003990/1 (COALESCE), of the Royal Society’s Global Challenges Research Fund Grant as well as of the European Research Council’s Advanced Fellow Grant QuantCom.}
	\thanks{J. An and L. Gan are with the School of Information and Communication Engineering, University of Electronic Science and Technology of China (UESTC), Chengdu, Sichuan 611731, China. L. Gan is also with the Yibin Institute of UESTC, Yibin, Sichuan 644000, China. (E-mail: jiancheng$\_$an@163.com; ganlu@uestc.edu.cn).}
	\thanks{C. Xu, Y. Liu and L. Hanzo are with the School of Electronics and Computer Science, University of Southampton, SO17 1BJ, UK. (E-mail: cx1g08@soton.ac.uk; yl6g15@ecs.soton.ac.uk; lh@ecs.soton.ac.uk).}
	\thanks{L. Wang is with Huawei Technology Sweden R\&D Competence Center, 16994 Stockholm, Sweden. (E-mail: leo.li.wang@huawei.com).}}

%\markboth{}%

\maketitle

\begin{abstract}
In Reconfigurable intelligent surface (RIS)-assisted systems the acquisition of channel state information (CSI) and the optimization of the reflecting coefficients constitute a pair of salient design issues. In this paper, a novel channel training protocol is proposed, which is capable of achieving a flexible performance vs. signalling and pilot overhead as well as implementation complexity trade-off. More specifically, \emph{first of all}, we conceive a holistic channel estimation protocol, which integrates the existing channel estimation techniques and passive beamforming design. \emph{Secondly}, we propose a new channel training framework. {\color{black}{In contrast to the conventional channel estimation arrangements, our new framework divides the training phase into several periods, where the superimposed end-to-end channel is estimated instead of separately estimating the direct BS-user channel and cascaded reflected BS-RIS-user channels.}} As a result, the reflecting coefficients of the RIS are optimized by comparing the objective function values over multiple training periods. Moreover, the theoretical performance of our channel training protocol is analyzed and compared to that under the optimal reflecting coefficients. In addition, the potential benefits of our channel training protocol in reducing the complexity, pilot overhead as well as signalling overhead are also detailed. \emph{Thirdly}, we derive the theoretical performance of channel estimation protocols and our channel training protocol in the presence of noise for a single-input-single-output (SISO) scenario, which provides useful insights into the impact of the noise on the overall RIS performance. \emph{Finally}, our numerical simulations characterize the performance of the proposed protocols and verify our theoretical analysis. In particular, the simulation results demonstrate that our channel training protocol is more competitive than the channel estimation protocol at low signal-to-noise ratios.
\end{abstract}


\IEEEpeerreviewmaketitle

\section{Introduction}\label{s1}
\IEEEPARstart{R}{econfigurable} intelligent surface (RIS) and its various variants are capable of beneficially ameliorating the wireless propagation environment \cite{wu2019towards, 2019arXiv191105326T, large_han_2019, beyond_hu_2018, wireless_basar_2019}. More specifically, they are constituted by a large number of reconfigurable passive reflecting elements, which are configured by a smart controller. As a result, each RIS component is configured to impose the required phase shift onto the incident signals, so that the received signals can be constructively enhanced at the intended user \cite{Cui2014, Renzo2019, 2020arXiv200703435L}, while simultaneously being nulled at the unintended user \cite{wu2019intelligent, wireless_basar_2019}. In contrast to conventional amplify/decode-and-forward cooperative communications, the RIS elements passively perform signal reflection without employing active radio frequency (RF) chains, hence eliminating the inevitable power consumption and delay of conventional delays \cite{bjornson2019intelligent, wu2019towards}. As a result, the RIS is capable of operating in a full-duplex mode without encountering the self-interference problem of complex relays \cite{Renzo2019, full_zhang_2015}. Therefore, the RIS is considered to be a promising candidate for creating favourable propagation environments for next-generation mobile communications \cite{a_saad_2020, wu2020intelligent, xu2019sixty}.

\begin{table*}
\centering
\caption{Contrasting our contribution to the state-of-the-art.}
\label{tab1}
\begin{tabular}{|c|c|c|c|c|c|c|c|c|}
\hline
\multirow{2}{*}{} & \multicolumn{4}{c|}{Channel Estimation Phase}                        & \multicolumn{4}{c|}{Joint Active and Passive Beamforming}         \\ \cline{2-9} 
                  & \begin{tabular}[c]{@{}c@{}}Reflected\\ Channels\end{tabular} & \begin{tabular}[c]{@{}c@{}}Direct\\ Channel\end{tabular} & \begin{tabular}[c]{@{}c@{}}RIS\\ Optimization\end{tabular} & \begin{tabular}[c]{@{}c@{}}MIMO\\ Setup\end{tabular} & \begin{tabular}[c]{@{}c@{}}Optimization\\ Criteria\end{tabular}    & \begin{tabular}[c]{@{}c@{}}MIMO\\ Setup\end{tabular}  & Methods   & \begin{tabular}[c]{@{}c@{}}Channel\\ Estimation\\  Errors\end{tabular}    \\ \hline
\cite{channel_mishra_2019}               &        \multicolumn{2}{c|}{ON/OFF method}           &       $\times$           &   MISO    &           \begin{tabular}[c]{@{}c@{}}Maximizing\\ the received power\end{tabular}    &  MISO     & \begin{tabular}[c]{@{}c@{}}Closed-form\\ solution\end{tabular}    &     $\times$                  \\ \hline
\cite{an_jensen_2020}               &        \multicolumn{2}{c|}{DFT-based method}           &       \checkmark           &   MISO    &           ---   &  ---     & ---   &     ---                  \\ \hline
\cite{you2020channel}               &         Grouped          &        Blocked        &       $\times$           &   MISO    &           ---   &  ---     & ---   &     ---                  \\ \hline
\cite{2019arXiv191203619C}               &    Sparsity                 &        Ignored        &        \checkmark          &  MU-MIMO     &        ---      &   ---     & ---  &         ---              \\ \hline
  \cite{wang2020channel}             &         Commonality          &  ON/OFF        &        $\times$         &  MU-MIMO      &        ---     &    ---   &  ---  &         ---             \\ \hline
\cite{zhang2019capacity}             &         ---           &      ---         &        ---          &  ---      &         \begin{tabular}[c]{@{}c@{}}Maximizing\\ the channel capacity\end{tabular}     &    \begin{tabular}[c]{@{}c@{}}point-to-point\\ MIMO\end{tabular}   &  AO  &         $\times$             \\ \hline
\cite{guo2019weighted}             &         ---           &      ---         &        ---          &  ---      &         \begin{tabular}[c]{@{}c@{}}Maximizing\\ the sum rate\end{tabular}     &    MU-MIMO   &  AO  &         $\times$             \\ \hline
\cite{cui2019secure}             &         ---           &      ---         &        ---          &  ---      &         \begin{tabular}[c]{@{}c@{}}Maximizing\\ the secrecy rate\end{tabular}     &    \begin{tabular}[c]{@{}c@{}}MISO with\\  an eavesdropper\end{tabular}   &  AO  &         $\times$             \\ \hline
  \cite{wu2019intelligent}             &         ---           &      ---         &        ---          &  ---      &         \begin{tabular}[c]{@{}c@{}}Minimizing\\ the transmit power\end{tabular}     &    MU-MIMO   &  SDR, AO  &         $\times$             \\ \hline
    {\color{black}{$\ast$}}            &                   \multicolumn{2}{c|}{DFT-based method}                &        \checkmark          &  MU-MIMO      &      \begin{tabular}[c]{@{}c@{}}Minimizing\\ the transmit power\end{tabular}        &   MU-MIMO   &  SDR, AO   &      {\color{black}{\checkmark}}        \\ \hline
                  & \multicolumn{4}{c|}{\color{black}{Channel Training Phase}}                          & \multicolumn{4}{c|}{Conventional Multi-user Downlink Beamforming} \\ \hline
     {\color{black}{$\star$}}              &                   \multicolumn{2}{c|}{Superimposed}                &        \checkmark          &    MU-MIMO   &            \begin{tabular}[c]{@{}c@{}}Minimizing\\ the transmit power\end{tabular}  &    MU-MIMO    &  SDR  &       {\color{black}{\checkmark}}                \\ \hline
\end{tabular}
\end{table*}

Nonetheless, some new challenges have arisen in RIS-assisted wireless systems. One of the major difficulties is the acquisition of the channel state information (CSI), which is particularly challenging in RIS-assisted systems due to the lack of any baseband signal processing capability at the RIS \cite{you2020channel, you2020intelligent, zheng2019intelligent, wang2020channel, AN_WCL_THE}. More explicitly, the downlink (DL) channels of the base station (BS)-RIS link and the RIS-user link cannot be estimated separately as in the conventional pilot-based approach. In order to solve this problem, sophisticated channel estimation methods have been proposed for estimating the cascaded BS-RIS and RIS-user channels \cite{channel_mishra_2019, an_jensen_2020, wang2020channel}. Specifically, in \cite{channel_mishra_2019}, a simple ON/OFF method is proposed, where only a single RIS element is switched on at each time slot (TS) and thus the corresponding reflected BS-RIS-user channel can be estimated without interference. Furthermore, the discrete Fourier transformation (DFT) matrix is applied in \cite{an_jensen_2020} for optimizing the reflecting coefficients (RCs) at the RIS during the channel estimation phase, which is capable of minimizing the mean square error (MSE) of the channel estimates. However, the method of either \cite{channel_mishra_2019} or \cite{an_jensen_2020} requires the same number of pilot symbols as that of the reflecting elements. To overcome this limitation, Wang \emph{et al.} \cite{wang2020channel} reduced the number of pilot symbols required for estimating the cascaded BS-RIS-user channels by exploiting the commonality of the BS-RIS link between multiple users \cite{wang2020channel}. More explicitly, a specific user's cascaded BS-RIS-user channel is first estimated, while muting all other users. Following this, the estimation of the reflected BS-RIS-user channels of other users is simplified by estimating the scaling factors relative to the specific user. {\color{black}{Inspired by \cite{wang2020channel}, the authors of \cite{2019arXiv191203619C} exploited the spatial sparsity of the BS-RIS link, so that the pilot overhead required for the specific user is further reduced. Moreover, the RIS elements are arranged into groups by You \emph{et al.} \cite{you2020channel}, where the combined reflected links are estimated by a reduced number of pilots at the cost of moderate channel estimation accuracy erosion.}} The reader might like to refer to \cite{zheng2019intelligent, you2020channel, wang2020channel} and the references therein for comprehensive surveys on the state-of-the-art channel estimation techniques designed for RIS-assisted systems.

On the other hand, the optimization of the RIS phase shifts is another major challenge in practical RIS-assisted systems \cite{guo2019weighted, zheng2019intelligent, zhou2020robust, nadeem2019intelligent, zhang2019capacity, zhou2020intelligent, abeywickrama2020intelligent, yang2020intelligent, TWC_PAN_MULTICELL, AN_TVT_optimal}. Recently, the RIS configuration of point-to-point multiple-input-multiple-output (MIMO) systems has been optimized in \cite{zhang2019capacity} in terms of maximizing the channel capacity. More specifically, in \cite{zhang2019capacity} the optimization of the RC and the transmit covariance matrix are performed alternately, where the optimal closed-form solution of each objective is derived under the condition that the other is determined. Furthermore, \cite{guo2019weighted} extends the contribution of \cite{zhang2019capacity} to the multiuser scenario by maximizing the sum rate of multiple users, where practical reflecting elements having discrete phase shifts are considered. In \cite{wu2019intelligent, wu2018intelligent}, Wu \emph{et al.} studied joint passive beamforming at the RIS and active transmit beamforming at the BS in multi-user MIMO scenarios, where the optimization objective is to minimize the total transmit power, subject to the independent signal-to-interference-plus-noise ratio (SINR) constraints for all users. The alternating optimization (AO) method based on semidefinite relaxation (SDR) is applied in order to obtain accurate approximate solutions for the RCs at the RIS and for the transmit beamforming vectors of multiple users. Additionally, the square law of the average received power vs. the number of reflecting elements is also summarized in \cite{wu2019intelligent}. As a further advance, the same authors extended the work of \cite{wu2019intelligent, wu2018intelligent} to the scenario of finite-precision phase shifts at the RIS \cite{wu2019beamforming, wu2019beamforming_1}. More recently, RISs have also been integrated with other existing technologies, such as physical-layer security \cite{chen2019intelligent, shen2019secrecy, cui2019secure, guan2019intelligent}, unmanned aerial vehicles (UAV) \cite{reconfigurable_li_2020}, non-orthogonal multiple access (NOMA) \cite{fu2019intelligent, yang2020intelligent_1}, index modulation (IM) \cite{Sarath_TWC_INT}, simultaneous wireless information and power transfer (SWIPT) \cite{pan2019intelligent, wu2019joint, wu2019weighted, pan2020intelligent} and mobile edge computing (MEC) \cite{bai2019latency}.

At the time of writing, substantial research efforts are invested into the conception of channel estimation techniques and phase shift optimization. However, the existing methods require a high pilot overhead and impose high estimation complexity. For example, the number of pilot symbols required for estimating the reflected BS-RIS-user channels is proportional to the number of reflecting elements \cite{you2020channel, wang2020channel}. The RC optimization of the RIS requires the joint consideration of the transmit beamforming at the BS and the passive beamforming at the RIS \cite{wu2019beamforming, zhang2019capacity}. These issues have to be addressed for the practical deployment of RISs. Moreover, there is a paucity of literature on the effect of channel estimation errors on the performance of the RIS. Against this background, in this paper, we propose a novel channel training protocol for RIS-assisted multi-user communications, which strikes a flexible trade-off between the performance attained, the implementation complexity imposed and the pilot overhead. In Table \ref{tab1}, we boldly and explicitly contrast our contributions to the existing solutions, where $\ast$ and $\star$ represent our proposed channel estimation protocol and channel training protocol, respectively\footnote{The channel estimation protocol is constituted by an amalgam of classic channel estimation schemes and RIS configuration methods, while our new channel training protocol will be detailed in Section \ref{s4}.}. More specifically, the novel contributions of this paper are summarized as follows:
\begin{itemize}
    \item First of all, we conceive a holistic channel estimation protocol, which intrinsically integrates the channel estimation methods of \cite{channel_mishra_2019, an_jensen_2020} and the passive beamforming design of \cite{wu2019intelligent, wu2018intelligent}. More specifically, we extend the DFT-based channel estimation method of \cite{an_jensen_2020} to multiuser scenarios, which was originally developed for single-user multiple-input-single-output (MISO) scenarios. Furthermore, we demonstrate the performance advantages of the DFT-based channel estimation method, especially for slow-fading channels having a long coherence time.
    \item Secondly, we propose a novel channel training protocol. {\color{black}{The proposed arrangement divides the training phase into several periods, where the superimposed end-to-end channel is estimated instead of separately estimating the direct BS-user and reflected BS-RIS-user channels.}} During each training period, the RCs of the RIS are configured through the proposed configuration method. As a result, the RIS phase shifts are optimized by comparing the objective function values over multiple training periods. Moreover, we introduce a pair of new RC configuration methods.
    \item Thirdly, the theoretical performance of the proposed channel training protocol is derived in terms of the average received power for a single-input-single-output (SISO) scenario, which is compared to that of the optimal RC. Both our theoretical analysis and simulation results demonstrate that our channel training protocol strikes an attractive performance vs. pilot overhead trade-off.
    \item Finally, the theoretical analysis of our channel estimation and channel training protocols is derived in the presence of noise and verified by numerical simulations.
\end{itemize}

The rest of this paper is structured as follows. Section \ref{s2} introduces the system model of our RIS-assisted wireless network. In Section \ref{s3} and \ref{s4}, we introduce our channel estimation and channel training protocols, respectively. Furthermore, the configuration of RC and the corresponding theoretical analysis of our channel training protocol are also addressed in Section \ref{s4}. In Section \ref{s5}, the effects of noise on both the channel estimation protocol and on the channel training protocol are analysed, while Section \ref{s6} provides numerical simulations of the proposed protocols. Finally, Section \ref{s7} offers our conclusions.

\emph{Notations:} We use upper (lower) bold face letters for representing matrices (column vectors); Scalars are denoted by italic letters; ${\left(  \cdot  \right)^T}$, ${\left(  \cdot  \right)^H}$ and ${\left(  \cdot  \right)^ * }$ represent transpose, Hermitian transpose and conjugate, respectively; $\text{diag}\left( {\bf{v}} \right)$ denotes a diagonal matrix with each diagonal element being the corresponding element in ${\bf{v}}$; $\left\|  \cdot  \right\|$ is the Frobenius norm of a complex vector, while $\left| \cdot \right|$ denotes the magnitude of a complex number; ${\mathbb{E}}\left\{  \cdot  \right\}$ stands for the expected value; $\otimes $ represents the Kronecker product; We denote the $N \times N$ identity matrix as ${{\bf{I}}_N}$; ${\bf{0}}$ and ${\bf{1}}$ denote an all-zero vector and an all-one vector, respectively, with appropriate dimensions. Furthermore, $\left\lfloor  \cdot  \right\rfloor$ and $\left \lceil \cdot  \right \rceil$ represent the floor and ceiling operation, respectively; The $\log \left(  \cdot  \right)$ represents the logarithmic function; $\Re \left ( z \right )$ and $\Im \left ( z \right )$ denote the real and imaginary part of a complex number $z$, respectively; ${{\bf{M}}_{a:b,c:d}}$ represents the elements of the $a \sim b$th rows and $c \sim d$th columns extracted from the matrix ${\bf{M}}$; ${{\bf{S}}^{ - 1}}$ denotes the inverse of the square matrix ${\bf{S}}$; The distribution of a circularly symmetric complex Gaussian random vector with mean vector ${\bf{v}}$ and covariance matrix ${\bf{\Sigma }}$ is denoted by $ \sim {\mathcal{CN}}\left( {{\bf{v}},{\bf{\Sigma }}} \right)$, where $ \sim $ stands for ``distributed as"; $ \sim {\mathcal{U}}\left( {a,b} \right)$ denotes the uniform distribution in the interval $\left( {a,b} \right)$; ${{\mathbb{C}}^{x \times y}}$ denotes the space of $x \times y$ complex-valued matrices; $\angle z$ denotes the phase of a complex number; $a!$ denotes the factorial of the non-negative integer $a$.
%%%%%%%%%%%%%%%%%%%%%%%%%%%%%%%%%%%%%%%%%%%%%%%%%%%%%%%%%%%%%%%%%% System model
\section{System Model}\label{s2}
\begin{figure}[!t]
	\centering
	\includegraphics[width=8.8cm]{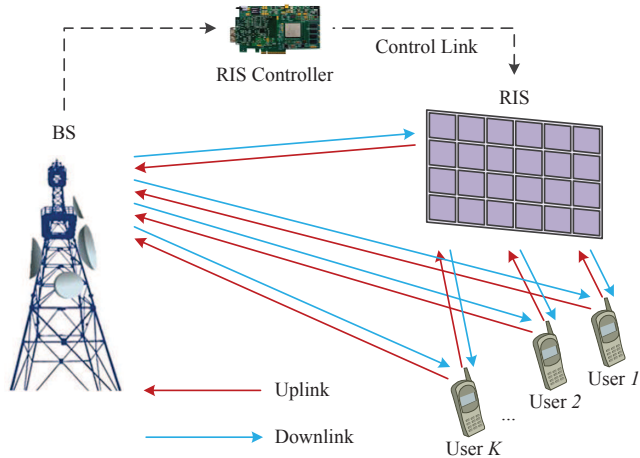}
	\caption{A RIS-assisted multiuser communication system.}
	\label{f1}
\end{figure}

Let us consider RIS-assisted multiuser communications in a single cell as shown in Fig. \ref{f1}, where a RIS is deployed to enhance the DL communications between a multi-antenna BS and $K$ single-antenna users. The number of transmit antennas (TAs) at the BS and that of the reflecting elements at the RIS are denoted by $M$ and $N$, respectively. The RIS is equipped with a smart controller that is capable of adjusting the RCs according to the real-time CSI \cite{Cui2014}. Additionally, the quasi-static flat-fading channel model is adopted for all links. In this paper, we consider a time-division duplexing (TDD) protocol for uplink (UL) as well as DL transmissions and assume the channel's reciprocity for the CSI acquisition in the DL based on the UL training.

The baseband equivalent DL channels spanning from the BS to RIS, from the RIS to user $k$, and from the BS to user $k$ are denoted by ${\bf{U}} \in {{\mathbb{C}}^{N \times M}}$, ${\bf{v}}_k^H \in {{\mathbb{C}}^{1 \times N}}$, and ${\bf{h}}_{d,k}^H \in {{\mathbb{C}}^{1 \times M}}$, respectively, with $k = 1,2, \cdots ,K$. Let $\boldsymbol{\varphi} =\left [ \varphi _{1},\varphi _{2},\cdots ,\varphi _{N} \right ]^{T}$ and define a diagonal matrix ${\boldsymbol{\Phi }} = diag\left( {{\varphi _1},{\varphi _2}, \cdots ,{\varphi _N}} \right)$ as the RC matrix of the RIS, where ${\varphi _n}$ denotes the RC of the $n$th element of the RIS, following $\left| {{\varphi _n}} \right| = 1$ for $n = 1,2, \cdots ,N$. In this paper, we assume that the phase shifts of ${\varphi _n}$ can be continuously varied in $\left[ {0,2\pi } \right)$, while in practice they are usually selected from several discrete values from $0$ to $2\pi $ for the sake of realistic implementation. The cascaded BS-RIS-user link is thus modelled as a concatenation of three components, namely, the BS-RIS links, RIS reflection with phase shifts, and RIS-user links. More specifically, we denote the reflected BS-RIS-user channel as $\mathbf{h}_{r,k}^{H}=\sum_{n=1}^{N}\mathbf{h}_{r,k,n}^{H}\varphi _{n}$, where $\mathbf{h}_{r,k,n}^{H}=\mathbf{v}_{k,n}^{H}\mathbf{U}_{n,:}$ denotes the reflected BS-RIS-user channel from the BS to user $k$ via the $n$th RIS element and $\mathbf{v}_{k,n}^{H}$ denotes the $n$th entry of $\mathbf{v}_{k}^{H}$.

We first consider the channel estimation in the UL phase. Regardless of the specific TS, the baseband signal ${\bf{y}}$ received at the BS is the sum of that via the direct BS-user link and reflected BS-RIS-user links, which can be expressed as
\begin{equation}\label{eq2-1}
{\bf{y}} = \sum\limits_{k = 1}^K {\left( {{{\bf{U}}^H}{{\bf{\Phi }}^H}{{\bf{v}}_k} + {{\bf{h}}_{d,k}}} \right){\sqrt \alpha  }{x_k}}  + {\bf{z}}
=\sum_{k=1}^{K}\mathbf{H}_{k}\boldsymbol{\varphi}'{\sqrt \alpha } x_{k}+\mathbf{z},
\end{equation}
where ${x_k}$ denotes the pilot symbol transmitted from user $k$ with zero mean and unit variance, $\alpha $ is the average power of the pilot symbols, which is assumed to be identical for all users. Furthermore, ${\bf{z}} \sim {\mathcal{CN}}\left( {{\bf{0}},\sigma _z^2{{\bf{I}}_M}} \right)$ denotes the additive white Gaussian noise (AWGN) at the BS, ${\boldsymbol{\varphi '}} = {\left[ {\begin{array}{*{20}{c}}
1&{{{\boldsymbol{\varphi }}^H}}
\end{array}} \right]^T}$ incorporates the phase shifts of the direct BS-user channel and reflected BS-RIS-user channels and ${{\bf{H}}_k} = \left[ {{{\bf{h}}_{d,k}},{{\bf{h}}_{r,k,1}},{{\bf{h}}_{r,k,2}}, \cdots ,{{\bf{h}}_{r,k,N}}} \right]$ incorporates the direct BS-user channel and reflected BS-RIS-user channels from the BS to user $k$, $k = 1,2, \cdots ,K$.

In the DL phase of data transmission, we use the linear transmit precoding (TPC) at the BS, where each user is assigned a dedicated beamforming vector. As a result, the complex baseband signal transmitted by the BS can be expressed as $\sum\nolimits_{k = 1}^K {{{\bf{w}}_k}{s_k}} $, where ${s_k}$ denotes the transmitted information symbol of user $k$ and ${{\bf{w}}_k}$ is the corresponding beamforming vector. It is assumed that ${s_k},k = 1,2, \cdots ,K$ are independent random variables with zero mean and unit variance. Therefore, the DL signal received by user $k$ from both the direct and reflected BS-RIS-user channels is then expressed as
\begin{equation}\label{eq2-2}
{r_k} = \left( {{\bf{v}}_{k}^H{\bf{\Phi }}{\bf{U}} + {\bf{h}}_{d,k}^H} \right)\sum\limits_{j = 1}^K {{{\bf{w}}_j}{s_j}}  + {n_k},\quad k = 1,2, \cdots ,K,
\end{equation}
where ${n_k}$ denotes the AWGN at the $k$th user's receiver. Without loss of generality, we assume the same noise floor level for all users, i.e., ${n_k} \sim {\mathcal{CN}}\left( {0,\sigma _n^2} \right),k = 1,2, \cdots ,K$.

%%%%%%%%%%%%%%%%%%%%%%%%%%%%%%%%%%%%%%%%%%%%%%%% Channel estimation and passive beamforming
\section{The Integration of Channel Estimation and Passive Beamforming}\label{s3}
In this section, we will introduce the existing RIS-assisted multiuser communication protocol relying on channel estimation. Specifically, the channel estimation methods of direct and reflected BS-RIS-user channels are introduced in Section \ref{s3-1}, while the optimization of the RCs at the RIS based on the estimated CSI is presented in Section \ref{s3-2}.
\subsection{Channel Estimation for RIS-assisted Multiuser Communication Systems}\label{s3-1}
Let us first consider two different channel estimation methods for RIS-assisted systems: the three-phase approach and the DFT-based approach. The three-phase method is proposed in \cite{wang2020channel}, while the DFT-based method is an extension of \cite{an_jensen_2020}, which was originally developed for the MISO scenario.
\subsubsection{Three-phase method}
The simplest CSI acquisition method of RIS-assisted communication systems is the ON/OFF method, which is based on turning each element of the RIS on and off, followed by estimating the direct BS-user channel and the reflected BS-RIS-user channels in sequence \cite{channel_mishra_2019}. Following a similar philosophy, the authors of \cite{wang2020channel} proposed a three-phase pilot-based channel estimation framework for multiuser scenarios. More specifically, in the first phase, all the reflecting elements are switched off for estimating the direct BS-user channel, i.e., ${{\bf{h}}_{d,k}},k = 1,2, \cdots ,K$. In the second phase, all the RIS reflecting elements are switched on, and merely one of the typical users, assumed to be user $1$, transmits non-zero pilot symbols to the BS. By leveraging the results of the first phase, the reflected BS-RIS-user channels of the specified user, i.e., $\mathbf{h}_{r,1,n},n=1,2,\cdots ,N$, can be estimated. Finally, in the third phase, merely user $2$ to user $K$ transmit their non-zero pilot symbols in the UL to the BS. In particular, since the reflected BS-RIS-user channels of $K$ users share the same BS-RIS link, the similar characteristics of the reflected BS-RIS-user channels between multiple users makes it possible to reduce the number of pilot symbols from $\left( {K - 1} \right)N$ to $\max \left( {K - 1,\left\lceil {{{\left( {K - 1} \right)N} \mathord{\left/
 {\vphantom {{\left( {K - 1} \right)N} M}} \right.
 \kern-\nulldelimiterspace} M}} \right\rceil } \right)$. The detailed procedures are given in \cite{wang2020channel}, which are omitted here for reasons of space-economy. In \cite{wang2020channel}, the authors verified that $\max \left( {K - 1,\left\lceil {{{\left( {K - 1} \right)N} \mathord{\left/
 {\vphantom {{\left( {K - 1} \right)N} M}} \right.
 \kern-\nulldelimiterspace} M}} \right\rceil } \right)$ pilot symbols provide sufficient degrees of freedom for estimating the reflected BS-RIS-user channels of users $2$ to $K$, i.e., $\mathbf{h}_{r,k,n},k=2,3,\cdots ,K,n=1,2,\cdots ,N$.

\begin{figure}[!t]
	\centering
	\includegraphics[width=8.8cm]{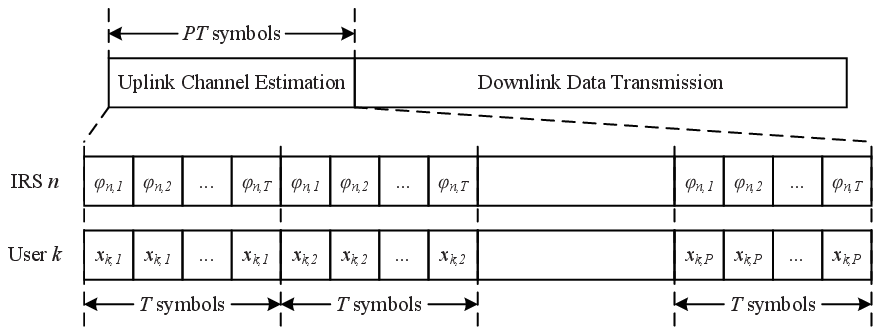}
	\caption{The frame structure of the channel estimation protocol.}
	\label{f2}
\end{figure}
\subsubsection{DFT-based method}
It may be readily seen that owing to reducing the number of pilot symbols, the three-phase method suffers from error propagation. To overcome this disadvantage, we conceive a generalised DFT-based channel estimation method, which is the extension of \cite{an_jensen_2020} initially designed for MISO scenarios. It is verified in \cite{an_jensen_2020} that the DFT-based RC optimization achieves the Cramér–Rao lower bound (CRLB) of the MSE of the channel estimation error.

The specific channel estimation protocol and its frame structure are shown in Fig. \ref{f2}, where the $PT$ pilot symbols are divided into $P$ groups. In each group, the same pilot vector is transmitted for $T$ TSs. The reflecting elements are always switched on throughout the channel estimation phase. More specifically, we have $P \ge K$ and $T \ge N + 1$ for estimating $N$ reflected BS-RIS-user channels and $1$ direct BS-user channel for $K$ users. In this paper, we use the minimum number of pilot symbols, i.e., $P = K$, $T = N + 1$. Therefore, for the $p$th pilot vector, i.e. the $p$th group, the UL received signal of the $t$th TS at the BS can be expressed as
\begin{equation}\label{eq3-1}
{{\bf{y}}_{p,t}} = \sum\limits_{k = 1}^K {{{\bf{H}}_k}{\bf{F}}_{:,t}^H{\sqrt \alpha }{x_{k,p}}}  + {{\bf{z}}_{p,t}},\quad p = 1,2, \cdots ,P,\quad t = 1,2, \cdots ,T,
\end{equation}
where ${\bf{F}}_{2:N + 1,t}^H$ denotes the RIS RC at the $t$th slot, and ${\bf{F}} \in {\mathbb{C}^{\left( {N + 1} \right) \times \left( {N + 1} \right)}}$ is the discrete Fourier transformation matrix.

Let ${{\bf{Y}}_p} = \left[ {{{\bf{y}}_{p,1}},{{\bf{y}}_{p,2}}, \cdots ,{{\bf{y}}_{p,N + 1}}} \right]$, ${{\bf{Z}}_p} = \left[ {{{\bf{z}}_{p,1}},{{\bf{z}}_{p,2}}, \cdots ,{{\bf{z}}_{p,N + 1}}} \right]$ and ${\bf{H}} = \left[ {{{\bf{H}}_1},{{\bf{H}}_2}, \cdots ,{{\bf{H}}_K}} \right]$. Upon collecting $\left( {N + 1} \right)$ TSs corresponding to the $p$th pilot vector, we arrive at:
\begin{equation}\label{eq3-2}
    {{\bf{Y}}_p} = {\bf{H}}\left( {{{\bf{I}}_K} \otimes {{\bf{F}}^H}} \right)\left( {\sqrt \alpha }{{{\bf{x}}_p} \otimes {{\bf{1}}_{N + 1,N + 1}}} \right) + {{\bf{Z}}_p},
\end{equation}
where ${{\bf{x}}_p} = {\left[ {{x_{1,p}},{x_{2,p}}, \cdots ,{x_{K,p}}} \right]^T}$ is the $p$th pilot vector.

Furthermore, let ${\bf{Y}} = \left[ {{{\bf{Y}}_1},{{\bf{Y}}_2}, \cdots ,{{\bf{Y}}_P}} \right]$ and ${\bf{Z}} = \left[ {{{\bf{Z}}_1},{{\bf{Z}}_2}, \cdots ,{{\bf{Z}}_P}} \right]$. After applying (\ref{eq3-2}) to all $P$ pilot vectors and simplifying, the result we get
\begin{equation}\label{eq3-3}
    {\bf{Y}} = {\bf{HG}} + {\bf{Z}},
\end{equation}
where ${\bf{G}} = \left( {{{\bf{I}}_K} \otimes {{\bf{F}}^H}} \right)\left( {\sqrt \alpha }{{\bf{X}} \otimes {{\bf{1}}_{N + 1,N + 1}}} \right)$ includes the effect of the RC at the RIS and of the pilot symbols transmitted in the UL by the users. The pilot design of ${\bf{X}}$ follows that of the conventional MIMO channel estimation \cite{training_biguesh_2006}.

According to (\ref{eq3-3}), the LS estimates of ${{\bf{\hat H}}}$ can be readily obtained by
\begin{equation}\label{eq3-4}
    {\bf{\hat H}} = {\bf{Y}}{{\bf{G}}^H}{\left( {{\bf{G}}{{\bf{G}}^H}} \right)^{ - 1}}.
\end{equation}

We note that the DFT-based channel estimation method requires at least $\left( {N + 1} \right)K$ pilot symbols, which is the same as that of the ON/OFF method. Nevertheless, the DFT-based channel estimation has a lower CSI estimation MSE than both the ON/OFF method and the three-phase method, which is crucial for the optimization of the RIS RC. We will demonstrate in Section \ref{s5} that in terms of achievable rate, the DFT-based channel estimation method remains the optimal option for slow-fading channels.

%%%%%%%%%%%%%%%%%%%%%%%%%%%%%%%%%%%%%%%%%%%%%%%%%%%%% Passive beamforming
\subsection{Joint Transmit Beamforming and RIS Configuration}\label{s3-2}
Upon obtaining the channel estimates of the direct and cascaded reflected BS-RIS-user links, the BS performs the multi-user TPC and the RIS carries out phase-shift configuration accordingly. Firstly, based on (\ref{eq2-2}), the SINR of user $k$ is given by
\begin{equation}\label{eq3-5}
SIN{R_k} = \frac{{{{\left| {\left( {{\bf{v}}_k^H{\bf{\Phi U}} + {\bf{h}}_{d,k}^H} \right){{\bf{w}}_k}} \right|}^2}}}{{\sum\nolimits_{j \ne k}^K {{{\left| {\left( {{\bf{v}}_{k}^H{\bf{\Phi U}} + {\bf{h}}_{d,k}^H} \right){{\bf{w}}_j}} \right|}^2}}  + \sigma _n^2}},\quad k = 1,2, \cdots ,K.
\end{equation}

Depending on the specific quality of service (QoS) requirements, the optimization criteria may differ, such as aiming for the maximized sum rate \cite{guo2019weighted}, minimized transmit power \cite{wu2019intelligent}, and maximized minimum SINR. In this paper, we opt for using the same optimization criteria as in \cite{wu2019intelligent}, which endeavours to minimize the total transmit power at the BS by jointly optimizing the TPC at the BS and RC at the RIS, subject to individual SINR constraints at all users. Therefore, the optimization problem can be formulated as
\begin{equation}\label{eq3-6}
\begin{array}{*{20}{l}}
{\mathop {\min }\limits_{{\bf{W}},{\boldsymbol{\varphi }}} }&{\sum\limits_{k = 1}^K {{{\left\| {{{\bf{w}}_k}} \right\|}^2}} }\\
{s.t.}&{\frac{{{{\left| {\left( {{\bf{v}}_k^H{\bf{\Phi U}} + {\bf{h}}_{d,k}^H} \right){{\bf{w}}_k}} \right|}^2}}}{{\sum\nolimits_{j \ne k}^K {{{\left| {\left( {{\bf{v}}_k^H{\bf{\Phi U}} + {\bf{h}}_{d,k}^H} \right){{\bf{w}}_j}} \right|}^2}}  + \sigma _n^2}} \ge {\gamma _k},\quad k = 1,2, \cdots ,K},\\
{}&{0 \le {\varphi _n} < 2\pi ,\quad n = 1,2, \cdots ,N},
\end{array}
\end{equation}
where ${\gamma _k} > 0$ is the minimum SINR requirement of user $k$. Solving (\ref{eq3-6}) is not trivial because of the non-convex constraints, since the optimization of the TPC and RC are coupled. In \cite{wu2019intelligent}, Wu \emph{et al.} discussed this problem in detail and proposed an alternating optimization algorithm based on SDR to obtain an accurate approximate solution.

In summary, we integrate the channel estimation technique with multi-user beamforming since there is a paucity of literature on the impact of channel estimation errors on the optimization of the RIS RC, this will be addressed in Section \ref{s5}. Moreover, the channel estimation protocol conceived in this section constitutes the baseline of our energy-efficient channel training-aided transmission protocol of Section \ref{s4}.
%%%%%%%%%%%%%%%%%%%%%%%%%%%%%%%%%%%%%%%%%%%%%%%%%%%%%%%%% channel training
\section{The Proposed Channel Training Framework}\label{s4}
In this section, we outline the proposed channel training scheme. Firstly, our channel training protocol is introduced in Section \ref{s4-1}. Furthermore, the specific training methods of the RIS RC are proposed in Section \ref{s4-2}, where the theoretical performance of our channel training framework is also analyzed in terms of the average received power at the users. Finally, Section \ref{s4-3} details the potential benefits of our channel training protocol over the existing channel estimation protocols in our RIS-enhanced wireless network.
%%%%%%%%%%%%%%%%%%%%%%%%%%%%%%%%%%%%%%%%%%%%%%%%%%%%%%%%% channel training protocol
\subsection{Channel Training Protocol}\label{s4-1}
\begin{figure}[!t]
	\centering
	\includegraphics[width=8.8cm]{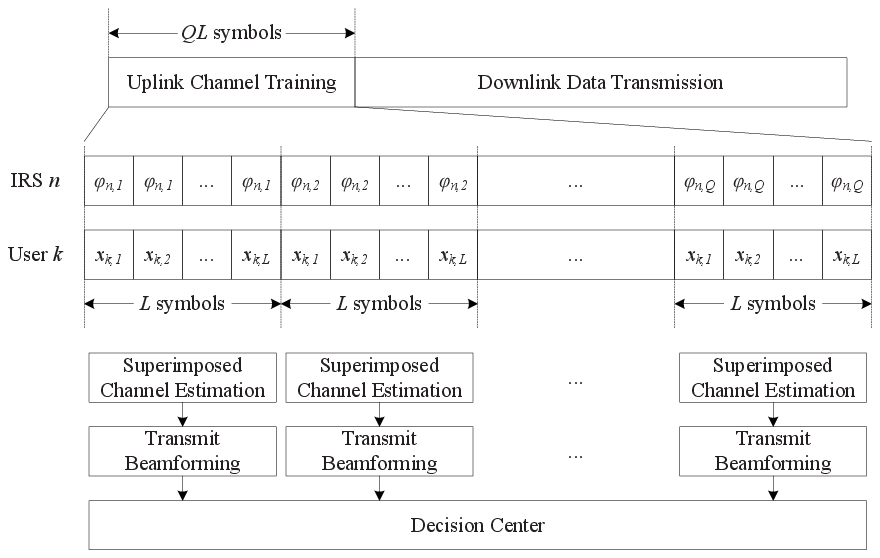}
	\caption{The channel training protocol and frame structure.}
	\label{f3}
\end{figure}

We first introduce our channel training protocol of Fig. \ref{f3}, which consists of two phases: the UL channel training and the DL data transmission, where $Q$ training periods of length $L$ are used for CSI acquisition and RIS configuration. In contrast to the existing channel estimation protocol of Section \ref{s3}, we suggest directly estimating the superimposed channel in each training period\footnote{{\color{black}{Note that we still have to perform channel estimation in our channel training protocol, but estimate the superimposed end-to-end channel (the amalgam of the direct BS-user channel and all reflected BS-RIS-user channels via the RIS) instead of a large number of reflected BS-RIS-user channels.}} The real meaning of `training' represents the training of RCs at the RIS by comparing the objective function values over multiple training periods.}. We adjust the RCs at the RIS over different training periods and repeat the superimposed channel estimation as well as the DL transmit beamforming design based on the instantaneous estimates of the current superimposed channel. Upon comparing the objective function values over multiple training periods, the optimal channel is selected from $Q$ candidates by the decision centre to assist in the DL data transmission. As a result, the phase shifts of the RIS reflecting elements and the corresponding DL TPC are also determined.

More specifically, we assume that the RC matrix of the $q$th training period is ${\boldsymbol{\Phi} _q},\ q = 1,2, \cdots ,Q$. Hence, the received signal of the $l$th TS of the $q$th training period can be expressed as
\begin{equation}\label{eq4-1}
    {{\bf{y}}_{q,l}} = \sum\limits_{k = 1}^K {\left( {{{\bf{U}}^H}{\bf{\Phi }}_q^H{{\bf{v}}_k} + {{\bf{h}}_{d,k}}} \right)\sqrt \alpha  {x_{k,l}}}  + {{\bf{z}}_{q,l}}= \sum\limits_{k = 1}^K {{{\bf{h}}_{q,k}}\sqrt \alpha {x_{k,l}}}  + {{\bf{z}}_{q,l}} = \sqrt \alpha  {{\bf{H}}_q}{{\bf{x}}_l} + {{\bf{z}}_{q,l}},
\end{equation}
where ${{\bf{h}}_{q,k}} = {{\bf{U}}^H}{\bf{\Phi }}_q^H{{\bf{v}}_k} + {{\bf{h}}_{d,k}}$ denotes the superimposed channel spanning from the $k$th user to the BS in the $q$th training period, ${x_{k,l}}$ is the pilot symbol transmitted from user $k$ in TS $l$, which is the same for all $Q$ training periods, while ${{\bf{H}}_q} = \left[ {{{\bf{h}}_{q,1}},{{\bf{h}}_{q,2}}, \cdots ,{{\bf{h}}_{q,K}}} \right]$ denotes the UL superimposed channel combining a single direct BS-user channel and $N$ reflected BS-RIS-user channels in the $q$th training period.

Upon collecting the signals of $L$ TSs in the $q$th training period, the UL signals received by the BS in the $q$th training period can be expressed as
\begin{equation}\label{eq4-2}
{{\bf{Y}}_q} = \left[ {{{\bf{y}}_{q,1}},{{\bf{y}}_{q,2}}, \cdots ,{{\bf{y}}_{q,L}}} \right] = \sqrt \alpha  {{\bf{H}}_q}{\bf{X}} + {{\bf{Z}}_q},
\end{equation}
where ${\bf{X}} = \left[ {{{\bf{x}}_1},{{\bf{x}}_2}, \cdots ,{{\bf{x}}_L}} \right]$ is the pilot matrix used for estimating the superimposed channel impinging from the $K$ users to the BS, which is identical for $Q$ training periods, while ${{\bf{Z}}_q} = \left[ {{{\bf{z}}_{q,1}},{{\bf{z}}_{q,2}}, \cdots ,{{\bf{z}}_{q,L}}} \right]$ denotes the noise matrix at the BS during the $q$th training period.

As a result, the estimation problem of the superimposed channel in the $q$th training period is equivalent to the conventional MIMO channel estimation, which has been widely studied for nearly three decades \cite{training_biguesh_2006, a_yin_2013}. In this paper, we harness the orthogonal pilot design of conventional MIMO channel estimation \cite{training_biguesh_2006}, where we have $L = K$. Following \cite{training_biguesh_2006}, the LS estimate of the superimposed channel ${{\bf{H}}_q}$ of the $q$th training period is given by
\begin{equation}\label{eq4-3}
{{{\bf{\hat H}}}_{q,LS}} = \frac{1}{{K\sqrt \alpha  }}{{\bf{Y}}_q}{{\bf{X}}^H}.
\end{equation}

After obtaining the estimates of the superimposed channel in the $q$th training period, the optimization problem of (\ref{eq3-6}) is reduced to
\begin{equation}\label{eq4-4}
\begin{array}{*{20}{l}}
{\mathop {\min }\limits_{{{\bf{W}}_q}} }&{\sum\limits_{k = 1}^K {{{\left\| {{{\bf{w}}_{q,k}}} \right\|}^2}} }\\
{s.t.}&{\frac{{{{\left| {{\bf{h}}_{q,k}^H{{\bf{w}}_{q,k}}} \right|}^2}}}{{\sum\nolimits_{j \ne k}^K {{{\left| {{\bf{h}}_{q,k}^H{{\bf{w}}_{q,j}}} \right|}^2}}  + \sigma _n^2}} \ge {\gamma _k},\forall k},
\end{array}
\end{equation}
where ${{\bf{W}}_q} = \left[ {{{\bf{w}}_{q,1}},{{\bf{w}}_{q,2}}, \cdots ,{{\bf{w}}_{q,K}}} \right]$ denotes the DL TPC matrix at the BS of the $q$th training period.

Note that (\ref{eq4-4}) represents the conventional power minimization problem of the multiuser DL MIMO channel. In \cite{optimal_bjornson_2014}, the general formula of the optimal TPC vectors is given by
\begin{equation}\label{eq4-5}
    {{\bf{\hat w}}_{q,k}} = \sqrt {{p_k}} \frac{{{{\left( {{{\bf{I}}_M} + \sum\nolimits_{j = 1}^K {\frac{{{\lambda _j}}}{{{\sigma_n ^2}}}{{\bf{h}}_{q,j}}{\bf{h}}_{q,j}^H} } \right)}^{ - 1}}{{\bf{h}}_{q,k}}}}{{\left\| {{{\left( {{{\bf{I}}_M} + \sum\nolimits_{j = 1}^K {\frac{{{\lambda _j}}}{{{\sigma_n ^2}}}{{\bf{h}}_{q,j}}{\bf{h}}_{q,j}^H} } \right)}^{ - 1}}{{\bf{h}}_{q,k}}} \right\|}},
\end{equation}
where ${{p_k}}$ denotes the TPC power for user $k$ while ${\lambda _j} \ge 0$ is the Lagrange multiplier associated with the $k$th SINR constraint \cite{optimal_bjornson_2014}. The $K$ unknown TPC powers can be obtained by solving the following linear equations:
\begin{equation}\label{eq4-6}
    \left[ {{p_1},{p_2}, \cdots ,{p_K}} \right] = {\sigma _n^2}{{\bf{1}}_{1 \times K}}{{\bf{M}}^{ - 1}},
\end{equation}
where the $\left( {i,j} \right)$th element of the matrix ${\bf{M}} \in \mathbb{R}{^{K \times K}}$ is given by
\begin{equation}\label{eq4-7}
    {\left[ {\bf{M}} \right]_{i,j}} = \left\{ {\begin{array}{*{20}{l}}
{\frac{1}{{{\gamma _i}}}{{\left| {{\bf{h}}_{q,i}^H{{{\bf{\tilde w}}}_{q,i}}} \right|}^2},}&{i = j},\\
{ - {{\left| {{\bf{h}}_{q,j}^H{{{\bf{\tilde w}}}_{q,i}}} \right|}^2},}&{i \ne j},
\end{array}} \right.
\end{equation}
and ${{{\bf{\tilde w}}}_{q,k}} = \frac{1}{{\sqrt {{p_k}} }}{{{\bf{\hat w}}}_{q,k}}$ is the TPC direction of user $k$. In addition, the Lagrange multipliers can be computed either by convex optimization \cite{an_luo_2006} or by the fixed-point iteration-based equations of \cite{linear_wiesel_2006, transmitter_yu_2007}.

By carrying out the superimposed channel estimation in (\ref{eq4-3}) and the DL TPC in (\ref{eq4-5}) for all $Q$ training periods, we obtain $Q$ candidate channels and their corresponding TPC matrix. Next, we proceed to select the optimal channel to assist the DL data transmission by continuing to adopt the criteria of minimizing the transmit power. More specifically, this optimization problem can be expressed as
\begin{equation}\label{eq4-8}
\begin{array}{*{20}{l}}
{\mathop {\min }\limits_q }&{\sum\limits_{k = 1}^K {{{\left\| {{{{\bf{\hat w}}}_{q,k}}} \right\|}^2}} }\\
{s.t.}&{1 \le q \le Q}.
\end{array}
\end{equation}
Once the index $\hat q$ of the optimal training period is obtained, the corresponding RC matrix, superimposed channel, and TPC matrix are determined accordingly by ${\boldsymbol{\hat \Phi }} = {{\boldsymbol{\Phi }}_{\hat q}}$, ${\bf{\hat H}} = {{\bf{\hat H}}_{\hat q}}$, ${\bf{\hat W}} = {{\bf{\hat W}}_{\hat q}}$.

%%%%%%%%%%%%%%%%%%%%%%%%%%%%%%%%%%%%%%%%%%%%%%%%%%%%%%%%%%%%% Training methods
\subsection{Configuration of RC at the RIS}\label{s4-2}
In this section, we consider a pair of training methods for the configuration of the RIS RC over $Q$ training periods, which provides useful insight into the performance characteristics of our channel training protocol.
\subsubsection{\textbf{Random configuration}}
Let us first consider the simple random configuration, namely, where the phase shifts of the RIS reflecting elements are randomly generated from a uniform distribution in each training period, which can be expressed as
\begin{equation}\label{eq4-9}
\angle {\varphi _{q,n}} \sim {\mathcal{U}}\left[ {0,2\pi } \right),\quad q = 1,2, \cdots ,Q,\quad n = 1,2, \cdots ,N.
\end{equation}

Next, we will analyze the scaling law of the average received power at the user for characterizing the performance of our channel training protocol relying on a random configuration. For the sake of simplicity, we assume that $M = 1$ and $K=1$ to gain essential insights. Let us first consider the case where the number of RIS reflecting elements is $1$, which will provide theoretical support for the more general scenario of $N$ elements. When the number of RIS reflecting elements is $1$, the average received power ${P_u}$ at the user employing our channel training protocol with random configuration is given by
\begin{equation}\label{eq4-10}
    {P_u} = P\mathbb{E}\left\{ {{{\left| {{h^*}} \right|}^2}} \right\} = P\mathbb{E}\left\{ {\mathop {\max }\limits_{q = 1,2, \cdots ,Q} \left\{ {\left| {{\varphi _q}h_r^* + h_d^*} \right|} \right\}} \right\},
\end{equation}
where $P$ is the transmit power at the BS, $h_r^* = u{v^*}$ denotes the cascaded reflected BS-RIS-user channel via the RIS without considering the phase shift, while $u$ and $v$ represent the degradation of ${\bf{U}}$ and ${{\bf{v}}_k}$, respectively. Since the BS and RIS are often on a tower or tall buildings, typically line-of-sight (LoS) propagation associated with only a few scatterers are observed. Hence the channel between the BS and RIS generally has a longer coherence time and a higher Rician factor than both the BS-user channel and the RIS-user channel. Therefore, it is reasonable to assume that ${h_r} \sim \mathcal{CN}\left( {0,\rho _r^2} \right)$, where $\rho _r^2$ is the path loss of the reflected BS-RIS-user channel via the RIS. Based on this assumption, the average received power of the user employing our channel training protocol with random configuration for the SISO scenario is summarized in \textbf{\emph{Proposition 1}}.

\textbf{\emph{Proposition 1:}} Upon assuming ${h_r} \sim \mathcal{CN}\left( {0,\rho _r^2} \right)$ and ${h_d} \sim \mathcal{CN}\left( {0,\rho _d^2} \right)$, we have:
\begin{equation}\label{eq4-11}
    {P_u} = P\left( {\rho _r^2 + \rho _d^2 + \frac{\pi }{2}{\rho _r}{\rho _d}g\left( Q \right)  } \right),
\end{equation}
where we have
\begin{equation}\label{eq4-12}
    g\left( Q \right) = \frac{{Q!}}{{{\pi ^Q}}}\sum\limits_{i = 1}^{\left\lceil {{{\left( {Q - 1} \right)} \mathord{\left/
 {\vphantom {{\left( {Q - 1} \right)} 2}} \right.
 \kern-\nulldelimiterspace} 2}} \right\rceil } {{{\left( { - 1} \right)}^{i + 1}}f\left( {Q - 2i} \right){\pi ^{Q - 2i}}},
\end{equation}
with $f\left( a \right) = 2$ for $a = 0$ and $f\left( a \right) = {1 \mathord{\left/
 {\vphantom {1 {a!}}} \right.
 \kern-\nulldelimiterspace} {a!}}$ for $a > 0$.

\emph{Proof:} {\color{black}{Please refer to Appendix A.}}

Next, let us consider the more general case of $N \ge 2$. The average received signal power can be expressed as
\begin{equation}\label{eq4-22}
        {P_u} = P{\mathbb{E}}\left\{ {\mathop {\max }\limits_{q = 1,2, \cdots ,Q} \left\{ {{{\left| {{h^H}} \right|}^2}} \right\}} \right\}= P{\mathbb{E}}\left\{ {\mathop {\max }\limits_{q = 1,2, \cdots ,Q} \left\{ {{{\left| {\sum\limits_{n = 1}^N {{\varphi _{q,n}}h_{r,n}^H}  + h_d^H} \right|}^2}} \right\}} \right\}.
\end{equation}

Since it is not as tractable to find the explicit solution of (\ref{eq4-22}) as in the case of $N = 1$, for $N \ge 2$, we provide an upper bound for characterizing its theoretical performance. More specifically, the upper bound of (\ref{eq4-22}) is summarized in \textbf{\emph{Proposition 2}}.

\textbf{\emph{Proposition 2:}} Upon assuming that ${h_{r,n}} \sim {\cal C}{\cal N}\left( {0,\rho _r^2} \right),n = 1,2, \cdots ,N$ and ${h_d} \sim \mathcal{CN}\left( {0,\rho _d^2} \right)$, we have:
\begin{equation}\label{eq4-23}
{P_u} \le P\left( {N\rho _r^2 + \rho _d^2 + \frac{\pi }{2}N{\rho _r}{\rho _d}g\left( Q \right) + \frac{\pi }{4}N\left( {N - 1} \right)\rho _r^2{g^2}\left( Q \right)} \right).
\end{equation}

\emph{Proof:} {\color{black}{Please refer to Appendix B.}}

\textbf{\emph{Propositions 1 $ \sim $ 2}} characterize the performance of a randomly configured channel training protocol for SISO scenarios, which offers the following useful observations. Firstly, let us gain some insight into the nature of the upper bound of (\ref{eq4-23}). The RHS of (\ref{eq4-23}) holds, when $N$ reflected BS-RIS-user channels are highly correlated. For example, considering a localized RIS-assisted scenario having a moderate number of tightly-placed reflecting elements (e.g., the antenna spacing is less than half wavelength), the upper bound of (\ref{eq4-23}) is achievable. On the other hand, assuming that all reflecting elements are independent of each other, if we do not limit the number of training periods, it takes at most $NQ$ training periods to approach the upper bound of (\ref{eq4-23}). One of the feasible ways is to switch on only a single reflecting element at a time and mute all the others for $Q$ training periods. The upper bound can also be achieved by doing so for $N$ RIS elements. It is not hard to see that the RHS of (\ref{eq4-23}) is also the lower bound for $NQ$ training periods.

Additionally, some asymptotic behavior can also be observed from \textbf{\emph{Propositions 1 $ \sim $ 2}}. For example, when $Q = 1$, we have ${P_{u,random}} = P\left( {N\rho _r^2 + \rho _d^2} \right)$, which is consistent with the conclusion drawn for a single random configuration in \cite{wu2019intelligent}. Upon increasing $Q$, ${P_u}$ will be gradually improved. In particular, as $Q \to \infty $, we have ${P_{u,optimal}} = P\left( {N\rho _r^2 + \rho _d^2 + \frac{\pi }{2}N{\rho _r}{\rho _d} + \frac{\pi }{4}N\left( {N - 1} \right)\rho _r^2} \right)$, which matches the average received power with optimal ${\bf{\Phi }}$ \cite{wu2019intelligent}. More generally, we summarize the relationship between the average received power using our channel training protocol relying on a random configuration and the theoretical optimal received power as \emph{Corollary 2-1}.

\emph{Corollary 2-1:} Let us assume that ${h_{r,n}} \sim {\cal C}{\cal N}\left( {0,\rho _r^2} \right),n = 1,2, \cdots ,N$ and ${h_d} \sim \mathcal{CN}\left( {0,\rho _d^2} \right)$. As $N \to \infty $, we have:
\begin{equation}\label{eq4-28}
\frac{{{P_u}}}{{{P_{u,optimal}}}} \to {g^2}\left( Q \right),
\end{equation}
where ${P_u} \to \frac{\pi }{4}{N^2}\rho _r^2{g^2}\left( Q \right)$ and ${P_{u,optimal}} \to \frac{\pi }{4}{N^2}\rho _r^2$ characterize the asymptotic values of ${{P_u}}$ and ${P_{u,optimal}}$, respectively, as $N \to \infty $.

It can be seen from \emph{Corollary 2-1} that, compared to the average received power associated with the optimal ${\bf{\Phi }}$, the randomly configured channel training protocol has a modest power loss of $\left[ {1 - {g^2}\left( Q \right)} \right]$, which can be compensated by increasing the number of training periods. On the other hand, \emph{Corollary 2-1} assumes the ideal case in the absence of channel estimation errors. When considering practical communication systems in the presence of channel estimation errors, our channel training protocol will become more competitive, which will be verified in Section \ref{s6}.

%%%%%%%%%%%%%%%%%%%%%%%%%%%%%%%%%%%%%%%%%%%%%%%%%%%%%%%%%%%%%%%%%%%%%%%%%% Equipartition configuration
\subsubsection{\textbf{Equi-partition configuration}}

\begin{figure}[!t]
	\centering
	\includegraphics[width=7cm]{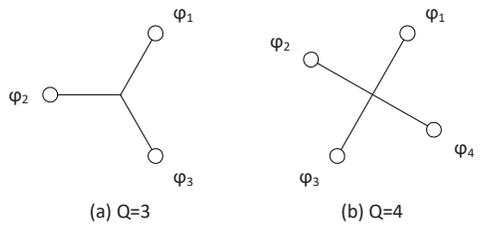}
	\caption{The equi-partition based configuration for $Q=3$ and $Q=4$, where we assume that $N=1$.}
	\label{f4}
\end{figure}

In our channel training protocol, we want to find the \emph{best} channel within a limited number of $Q$ training periods. It is clear that the random configuration does not utilize the $Q$ training periods effectively. Therefore, we propose a more sophisticated technique for the configuration of the phase shifts, which is based on the following philosophy: The more different RCs exist, the more different channels can be generated, thus leading to $Q$ candidate sets associated with more significant differences. In this paper, we use the Euclidean distance between the RC to quantify this difference. Therefore, the optimization problem of RC sets can be expressed as
\begin{equation}\label{eq4-29}
    \begin{array}{*{20}{l}}
{\mathop {\max }\limits_{{{\boldsymbol{\varphi }}_1},{{\boldsymbol{\varphi }}_2}, \cdots ,{{\boldsymbol{\varphi }}_Q}} }&{\sum\limits_{q = 1}^Q {\sum\limits_{q' = 1,q' \ne q}^Q {{{\left\| {{{\boldsymbol{\varphi }}_q} - {{\boldsymbol{\varphi }}_{q'}}} \right\|}^2}} } }.
\end{array}
\end{equation}

For an arbitrary one of the reflecting elements, the problem of (\ref{eq4-29}) is reduced to
\begin{equation}\label{eq4-30}
\begin{split}
\left[ {{\varphi _{1,n}},{\varphi _{2,n}}, \cdots ,{\varphi _{Q,n}}} \right]&= \mathop {\max }\limits_{{\varphi _{1,n}},{\varphi _{2,n}}, \cdots ,{\varphi _{Q,n}}} \sum\limits_{q = 1}^Q {\sum\limits_{q' = 1,q' \ne q}^Q {\sum\limits_{n = 1}^N {{{\left| {{\varphi _{q,n}} - {\varphi _{q',n}}} \right|}^2}} } } \\
 &= \mathop {\max }\limits_{{\varphi _{1,n}},{\varphi _{2,n}}, \cdots ,{\varphi _{Q,n}}} \sum\limits_{q = 1}^Q {\sum\limits_{q' = 1,q' \ne q}^Q {{{\left| {{\varphi _{q,n}} - {\varphi _{q',n}}} \right|}^2}} }.
\end{split}
\end{equation}

Furthermore, finding the maximum value of (\ref{eq4-30}) is equivalent to the constraint of zero on its partial derivative, which can be expressed as
\begin{equation}\label{eq4-31}
\sum\limits_{q' = 1}^Q {\Im \left( {{\varphi _{q',n}}\varphi _{q,n}^*} \right)}  = 0,q = 1,2, \cdots ,Q.
\end{equation}

It is not difficult to verify that the configuration of the phase shifts satisfying (\ref{eq4-31}) forms a regular $Q$-polygon, and one of the possible solutions can be expressed as
\begin{equation}\label{eq4-32}
{\varphi _{q,n}} = {\varphi _n}{e^{j\frac{{2\left( {q - 1} \right)\pi }}{Q}}},q = 1,2, \cdots ,Q,
\end{equation}
where ${\varphi _n} \sim {\mathcal{U}}\left[ {0,2\pi } \right)$ denotes the random initial phase. Fig. \ref{f4} gives examples for $Q=3$ and $Q=4$, where we assume $N=1$ for the sake of illustration.

Following a similar line of the proof followed for \textbf{\emph{Propositions 1 $\sim$ 2}}, the average received power of our channel training protocol associated with our equi-partition configuration is summarized in \textbf{\emph{Proposition 3}}.

\textbf{\emph{Proposition 3:}} Upon assuming ${h_{r,n}} \sim {\cal C}{\cal N}\left( {0,\rho _r^2} \right),n = 1,2, \cdots ,N$ and ${h_d} \sim \mathcal{CN}\left( {0,\rho _d^2} \right)$, we have:
\begin{equation}\label{eq4-33}
{P_u} \le P\left( {N\rho _r^2 + \rho _d^2 + \frac{\pi }{2}N{\rho _r}{\rho _d}\frac{{\sin \left( {{\pi  \mathord{\left/
 {\vphantom {\pi  Q}} \right.
 \kern-\nulldelimiterspace} Q}} \right)}}{{\left( {{\pi  \mathord{\left/
 {\vphantom {\pi  Q}} \right.
 \kern-\nulldelimiterspace} Q}} \right)}} + \frac{\pi }{4}N\left( {N - 1} \right)\rho _r^2\frac{{{{\sin }^2}\left( {{\pi  \mathord{\left/
 {\vphantom {\pi  Q}} \right.
 \kern-\nulldelimiterspace} Q}} \right)}}{{{{\left( {{\pi  \mathord{\left/
 {\vphantom {\pi  Q}} \right.
 \kern-\nulldelimiterspace} Q}} \right)}^2}}}} \right).
\end{equation}

\emph{Proof:} When the equi-partition configuration scheme of Fig. \ref{f4} is adopted, $\mathbb{E}\left\{ {\mathop {\max }\limits_{q = 1,2, \cdots ,Q} \left\{ {\cos {\theta _q}} \right\}} \right\}$ can be calculated as
\begin{equation}\label{eq4-34}
\mathbb{E}\left\{ {\mathop {\max }\limits_{q = 1,2, \cdots ,Q} \left\{ {\cos {\theta _q}} \right\}} \right\} = \frac{Q}{{2\pi }}\int\limits_{ - {\pi  \mathord{\left/
 {\vphantom {\pi  Q}} \right.
 \kern-\nulldelimiterspace} Q}}^{{\pi  \mathord{\left/
 {\vphantom {\pi  Q}} \right.
 \kern-\nulldelimiterspace} Q}} {\cos {\theta _1}d{\theta _1}}  = \frac{{\sin \left( {{\pi  \mathord{\left/
 {\vphantom {\pi  Q}} \right.
 \kern-\nulldelimiterspace} Q}} \right)}}{{\left( {{\pi  \mathord{\left/
 {\vphantom {\pi  Q}} \right.
 \kern-\nulldelimiterspace} Q}} \right)}}.
\end{equation}

{\color{black}{Upon substituting (\ref{eq4-34}) into (\ref{eq4-22}) and following the same considerations as Appendix B, we complete the proof.}} $\hfill\blacksquare$

It can be seen from \textbf{\emph{Proposition 3}} that the average received power of our channel training protocol using our equi-partition configuration has the same trend as that of the random configuration, but suffers from a smaller power loss of $\left( {1 - \frac{{{{\sin }^2}\left( {{\pi  \mathord{\left/
 {\vphantom {\pi  Q}} \right.
 \kern-\nulldelimiterspace} Q}} \right)}}{{{{\left( {{\pi  \mathord{\left/
 {\vphantom {\pi  Q}} \right.
 \kern-\nulldelimiterspace} Q}} \right)}^2}}}} \right)$ than the random configuration\footnote{Here we directly use the conclusion $\frac{{\sin \left( {{\pi  \mathord{\left/
 {\vphantom {\pi  Q}} \right.
 \kern-\nulldelimiterspace} Q}} \right)}}{{\left( {{\pi  \mathord{\left/
 {\vphantom {\pi  Q}} \right.
 \kern-\nulldelimiterspace} Q}} \right)}}  \ge  g\left( Q \right)$ without proof, which will be verified by our simulations.}. Nevertheless, the equi-partition configuration has some limitations in practical implementations, which will be detailed in Section \ref{s6}.

\begin{table*}
\centering
\caption{Contrasting our channel training protocol to the existing channel estimation protocols}
\label{tab2}
\begin{tabular}{|l||p{6cm}|p{6cm}|}
\hline
Type                   & Existing channel estimation protocols \cite{you2020channel, an_jensen_2020, wang2020channel, channel_mishra_2019, 2019arXiv191203619C, zhang2019capacity, guo2019weighted, cui2019secure, you2020intelligent, zheng2019intelligent}        & Our channel training protocol      \\ \hline\hline
System design             & Complex                & Simple          \\ \hline
Pilot overhead            & Large, $\left( {N + 1} \right)K$                & Small, $QK$     \\ \hline
Error propagation            & Severe                &   Mild   \\ \hline
Signalling overhead       & Large, $N{\log _2}B$      & Small, $\left\lceil {{{\log }_2}Q} \right\rceil $    \\ \hline
Channel model    &          Dependent                  &         Independent        \\ \hline
Performance    &          Better                  &         Slight performance penalty        \\ \hline
Optimization problems & \begin{tabular}[c]{@{}l@{}}Channel estimation protocol\\ Joint pilot design and RIS configuration\\ Joint active and passive beamforming\end{tabular} & Channel training protocol \\ \hline
\end{tabular}
\end{table*}
%%%%%%%%%%%%%%%%%%%%%%%%%%%%%%%%%%%%%%%%%%%%%%%%%%%%%%%%%%%%%%% Training benefits
\subsection{Benefits of Channel Training Protocol}\label{s4-3}
In this subsection, we will elaborate on the benefits of our superimposed channel training scheme compared to its existing separate channel estimation counterparts, based on the following five aspects:
\subsubsection{\textbf{Simplified system design}}
In contrast to the existing channel estimation protocols \cite{you2020channel, an_jensen_2020, wang2020channel, channel_mishra_2019, 2019arXiv191203619C, zhang2019capacity, guo2019weighted, cui2019secure, you2020intelligent, zheng2019intelligent}, the proposed channel training scheme simplifies the complex signal processing of the existing solutions. More specifically, the channel estimation of the RIS-assisted systems has to consider the joint design of the pilot symbols transmitted in the UL from the users and RC of the RIS, as well as the specific estimation protocol employed. Furthermore, in the data transmission phase, the RC and TPC have also to be optimized jointly, which is complex to implement. For example, in \cite{wang2020channel}, the authors considered a three-phase channel estimation protocol, which estimates the direct BS-user channel link, the reflected BS-RIS-user channels of a specific user via the RIS, and other users in the three phases. Moreover, Wu \emph{et al.} \cite{wu2019intelligent, wu2018intelligent} considered the joint design of the RC and the TPC, and solved the problem by the SDR convex optimization algorithm and alternate optimization. By contrast, the proposed channel training protocol is more concise. During each training period, we only have to estimate the superimposed MIMO channel and perform TPC without considering the influence of the RIS, which have been richly studied in conventional systems, leading to a lot of sophisticated algorithms \cite{training_biguesh_2006, a_yin_2013}. The configuration of the RC is completed over multiple training periods by a pre-established selection criterion, which effectively avoids the aforementioned problems and significantly simplifies the system design.
\subsubsection{\textbf{Flexible pilot overhead}}
Compared to the existing channel estimation protocols \cite{you2020channel, an_jensen_2020, wang2020channel, channel_mishra_2019, 2019arXiv191203619C, zhang2019capacity, guo2019weighted, cui2019secure, you2020intelligent, zheng2019intelligent}, the pilot overhead of our channel training protocol is more moderate, striking a flexible performance vs. overhead trade-off. More specifically, for the channel estimation protocol of \cite{channel_mishra_2019}, the pilot overhead of the simplest ON/OFF method mentioned is $\left( {N + 1} \right)K$. Although some estimation strategies have been designed for reducing the pilot overhead, they often introduce severe error propagation or sacrifice the performance. For example, \cite{wang2020channel} reduces the pilot overhead to $K + N + \max \left( {K - 1,\left\lceil {{{\left( {K - 1} \right)N} \mathord{\left/
 {\vphantom {{\left( {K - 1} \right)N} M}} \right.
 \kern-\nulldelimiterspace} M}} \right\rceil } \right)$ by taking advantage of the similarity of the BS-RIS link of the reflected BS-RIS-user channels between multiple users. On the other hand, \cite{you2020channel} reduces the pilot overhead to $\left( {D + 1} \right)K$ by dividing the RIS reflecting elements into $D$ groups, but suffers from a performance penalty. Nevertheless, the pilot overhead of channel estimation is still dependent on the number of reflecting elements $N$. By contrast, our channel training protocol avoids this pilot overhead of $QK$, thus striking a flexible performance vs. overhead trade-off.
\subsubsection{\textbf{Mitigated error propagation}}
Several channel estimation methods attempted to reduce the pilot overhead of the channel estimation \cite{you2020channel, wang2020channel, channel_mishra_2019}. However, these methods were designed for the sole purpose of reducing the pilot overhead, while ignoring the error propagation effects during the channel estimation phase. For example, in order to exploit the similarities of the reflected BS-RIS-user channels, an ON/OFF estimation strategy is used in \cite{wang2020channel}. Specifically, the RIS reflecting elements are muted first for estimating the direct BS-user channel, and then the direct BS-user channel is used for estimating the reflected BS-RIS-user channel. Since the estimated direct BS-user channel is inaccurate, it exacerbates the estimation error of the reflected BS-RIS-user channels. Additionally, \cite{wang2020channel} first estimates the reflected BS-RIS-user channels of a specific user, and then estimates that of all other users, which also precipitates the error propagation between different users. Moreover, the channel estimation errors can also affect the RC, thereby eroding the expected performance improvements attained by the RIS. By contrast, in the proposed channel training protocol, the channel estimation errors only affect the index of the training period deemed to be the best. Compared to channel estimation, the error propagation within our channel training protocol is significantly reduced. We will show in our subsequent numerical simulations and theoretical analysis that in the presence of channel estimation errors, bespoke channel training can even mitigate the performance gap imposed by the channel estimation protocol, despite its reduced system complexity and frugal pilot overhead.
\subsubsection{\textbf{Reduced signalling overhead}}
In the channel estimation protocol of \cite{you2020channel, an_jensen_2020, wang2020channel, channel_mishra_2019, 2019arXiv191203619C, zhang2019capacity, guo2019weighted, cui2019secure, you2020intelligent, zheng2019intelligent}, the RCs are optimized by the controller centre and then they have to be transmitted through the control link to the RIS, which requires the extra signalling of $N{\log _2}B$ bits to the RIS. Note that here we consider the practical implementation of the reflecting elements using finite-precision phase shifts, where $B$ represents the number of quantization order. The total number of bits required for control signalling is linearly proportional to the number of RIS elements and logarithmically to $B$. Fortunately, for our channel training protocol, the RIS configuration can be completed by only feeding back the index of the optimal training period to the RIS, hence the number of control signalling bits is $\left\lceil {{{\log }_2}Q} \right\rceil $. Hence, it is apparent that our channel training protocol significantly reduces the signalling overhead and thus also the transmission delay between the control centre and the RIS compared to the channel estimation protocol of \cite{you2020channel, an_jensen_2020, wang2020channel, channel_mishra_2019, 2019arXiv191203619C, zhang2019capacity, guo2019weighted, cui2019secure, you2020intelligent, zheng2019intelligent}.
\subsubsection{\textbf{Model independent}}
Compared to the channel estimation protocol of \cite{you2020channel, an_jensen_2020, wang2020channel, channel_mishra_2019, 2019arXiv191203619C, zhang2019capacity, guo2019weighted, cui2019secure, you2020intelligent, zheng2019intelligent}, our channel training method does not depend on the specific channel model via the RIS. At each training period, only the RCs have to be configured and the BS will perform the estimation of the superimposed channel. By contrast, in the existing schemes, the cascaded reflected BS-RIS-user channels are estimated, and the channel estimation algorithms generally have to consider, for example, the correlation between reflecting elements \cite{you2020channel}, or the similarity of the BS-RIS link between multiple users \cite{wang2020channel} or alternatively the sparsity of the BS-RIS link \cite{2019arXiv191203619C}. By contrast, our channel training method reduces the dependence on the specifics of the channel model and thus has a wider range of applications than the channel estimation protocol of \cite{you2020channel, an_jensen_2020, wang2020channel, channel_mishra_2019, 2019arXiv191203619C, zhang2019capacity, guo2019weighted, cui2019secure, you2020intelligent, zheng2019intelligent}.

In conclusion, our channel training scheme and the channel estimation protocols of \cite{you2020channel, an_jensen_2020, wang2020channel, channel_mishra_2019, 2019arXiv191203619C, zhang2019capacity, guo2019weighted, cui2019secure, you2020intelligent, zheng2019intelligent} are boldly contrasted in Table \ref{tab2}.

%%%%%%%%%%%%%%%%%%%%%%%%%%%%%%%%%%%%%%%%%%%%%%%%%%%%%%% Channel estimation errors
\section{The Effect of Channel Estimation Errors}\label{s5}
In this section, we evaluate the robustness of the channel estimation protocols of \cite{you2020channel, an_jensen_2020, wang2020channel, channel_mishra_2019, 2019arXiv191203619C, zhang2019capacity, guo2019weighted, cui2019secure, you2020intelligent, zheng2019intelligent} and our channel training protocol in the face of channel estimation errors. Following the same considerations as in Section \ref{s4-2}, we continue to use the average received power to characterize the effect of channel estimation errors. In this context, we consider the SISO scenario, where only a single user is served by a BS having a single TA.
\subsection{Theoretical Analysis of the Channel Estimation Protocol \cite{you2020channel, an_jensen_2020, wang2020channel, channel_mishra_2019, 2019arXiv191203619C, zhang2019capacity, guo2019weighted, cui2019secure, you2020intelligent, zheng2019intelligent}}\label{s5-1}
First, let us consider the channel estimation protocol of Section \ref{s3}. We assume that the estimated reflected BS-RIS-user channel coefficients and direct BS-user channel coefficients are
\begin{equation}\label{eq5-1}
    {\hat h_d} = {h_d} + {\varepsilon _d},\quad {\hat h_{r,n}} = {h_{r,n}} + {\varepsilon _{r,n}},n = 1,2, \cdots ,N,
\end{equation}
respectively, where ${\varepsilon _d} \sim {\mathcal{CN}}\left( {0,\sigma _d^2} \right)$ and ${\varepsilon _{r,n}} \sim {\mathcal{CN}}\left( {0,\sigma _r^2} \right),n = 1,2, \cdots ,N$ represent the estimation error of the direct BS-user channel and reflected BS-RIS-user channels via the $n$th RIS element, respectively.

Based on \cite{wu2018intelligent}, the RIS RC based on the channel estimates can be expressed as
\begin{equation}\label{eq5-2}
{\hat \varphi _n} = \frac{{\hat h_{r,n}^*{{\hat h}_d}}}{{\left| {\hat h_{r,n}^*{{\hat h}_d}} \right|}},n = 1,2, \cdots ,N.
\end{equation}

Due to the influence of channel estimation errors, the rotated reflected BS-RIS-user channels cannot be perfectly coherently-superimposed on the direct BS-user channel at the receiver, which will result in some power loss. More specifically, the average power received in the presence of channel estimation errors is summarized in \textbf{\emph{Proposition 4}}.

\textbf{\emph{Proposition 4:}} Upon assuming ${h_{r,n}} \sim {\cal C}{\cal N}\left( {0,\rho _r^2} \right),n = 1,2, \cdots ,N$, ${h_d} \sim \mathcal{CN}\left( {0,\rho _d^2} \right)$, ${\varepsilon _d} \sim {\mathcal{CN}}\left( {0,\sigma _d^2} \right)$ and ${\varepsilon _{r,n}} \sim {\mathcal{CN}}\left( {0,\sigma _r^2} \right),n = 1,2, \cdots ,N$, we have:
\begin{equation}\label{eq5-3}
{P_u} = P\left( {N\rho _r^2 + \rho _d^2 + \frac{{\pi N\rho _r^2\rho _d^2}}{{2\sqrt {\left( {\rho _r^2 + \sigma _r^2} \right)\left( {\rho _d^2 + \sigma _d^2} \right)} }} + \frac{{\pi N\left( {N - 1} \right)\rho _r^4}}{{4\left( {\rho _r^2 + \sigma _r^2} \right)}}} \right).
\end{equation}

\emph{Proof:} {\color{black}{Please refer to Appendix C.}}

\textbf{\emph{Proposition 4}} quantifies the influence of channel estimation errors on the average received power of the user. It can be seen that when the channel estimation errors are severe, we have ${P_u} = P\left( {N\rho _r^2 + \rho _d^2} \right)$, which is consistent with the conclusion of single random configuration. Moreover, ${P_u}$ is equivalent to the optimal received power when there is no channel estimation error. Additionally, as $N$ increases, the asymptotic behaviour of average received power in the face of channel estimation errors is described in \emph{Corollary 4-1}.

\emph{Corollary 4-1:} Upon assuming ${\varepsilon _d} \sim {\mathcal{CN}}\left( {0,\sigma _d^2} \right)$ and ${\varepsilon _{r,n}} \sim {\mathcal{CN}}\left( {0,\sigma _r^2} \right),n = 1,2, \cdots ,N$. As $N \to \infty $, we have
\begin{equation}\label{eq5-11}
\frac{{{P_u}}}{{{P_{u,optimal}}}} \to \frac{{\rho _r^2}}{{\rho _r^2 + \sigma _r^2}}.
\end{equation}

\emph{Corollary 4-1} shows that with the increase of $N$, the average received power is determined by the path loss and by the estimation error of the reflected BS-RIS-user channels. This also implies that accurately estimating the reflected BS-RIS-user channels is beneficial for improving the performance of RIS equipped with numerous reflecting elements.

\begin{table*}[]
\centering
\caption{The comparison of three channel estimation methods}
\label{tab3}
\begin{tabular}{|l|l|l|l|l|}
\hline
  & Pilot overhead & Estimation accuracy & Applicable scenario     & Achievable rate \\ \hline
ON/OFF method \cite{channel_mishra_2019}& Large          & Low                & Slow fading channel & Low             \\ \hline
Three-phase method \cite{wang2020channel} &   Middle      &     Middle    &   Fast fading channel     & Middle       \\ \hline
DFT-based method \cite{an_jensen_2020}&      Small &      High   &      Slow fading channel            & High            \\ \hline
\end{tabular}
\end{table*}

Next, we will evaluate the performance of three channel estimation algorithms in terms of their achievable rate under their respective channel estimates. The ergodic achievable rate in terms of bits per second per Hertz (b/s/Hz) is approximately given by
\begin{equation}\label{eq5-12}
R = \frac{{{\tau _{co}} - \tau }}{{\tau _{co}}}{\log _2}\left( {1 + \frac{{P\mathbb{E}\left\{ {{{\left| {\sum\nolimits_{n = 1}^N {{{\hat \varphi }_n}h_{r,n}^H}  + h_d^H} \right|}^2}} \right\}}}{{{\sigma ^2}}}} \right),
\end{equation}
where $\tau $ is the number of symbols required for performing channel estimation, while ${\tau _{co}}$ is the number of symbols within the coherence time.

\subsubsection{The ON/OFF method}
We first consider the simplest ON/OFF method described in \cite{channel_mishra_2019}, where the reflecting elements are activated one by one to estimate the direct BS-user channel and the reflected BS-RIS-user channel. As demonstrated in \cite{channel_mishra_2019}, the minimum number of pilot symbols required for estimating $N$ reflected BS-RIS-user channels plus a direct BS-user channel and the MSE of the channel estimates are
\begin{equation}\label{eq5-13}
\tau \left| {_{ON/OFF}} \right. = N + 1,\quad \sigma _d^2\left| {_{ON/OFF}} \right. = {\sigma ^2},\quad \sigma _{r,n}^2\left| {_{ON/OFF}} \right. = 2{\sigma ^2},n = 1,2, \cdots ,N,
\end{equation}
respectively, where we have ${\sigma ^2} = {{\sigma _z^2} \mathord{\left/
 {\vphantom {{\sigma _z^2} \alpha }} \right.
 \kern-\nulldelimiterspace} \alpha }$.
\subsubsection{The three-phase method}
When considering a single user, the direct BS-user channel estimation of the three-phase method of \cite{wang2020channel} degenerates into the ON/OFF estimation strategy of \cite{channel_mishra_2019}. First, all the reflecting elements are switched off for estimating the direct BS-user channel. Then all the reflecting elements are switched on to complete the estimation of the remaining reflected BS-RIS-user channels and the RCs are optimized by the DFT matrix. As described in \cite{wang2020channel}, the required number of pilot symbols and the MSE of channel estimates are given by
\begin{equation}\label{eq5-14}
    \tau \left| {_{three - phase}} \right. = N + 1,\quad \sigma _d^2\left| {_{three - phase}} \right. = {\sigma ^2},\quad \sigma _{r,n}^2\left| {_{three - phase}} \right. = \frac{{2{\sigma ^2}}}{N},n = 1,2, \cdots ,N.
\end{equation}

\subsubsection{The DFT-based method}
Finally, let us consider the optimal channel estimation scheme, i.e. the DFT-based method of \cite{an_jensen_2020}. The reflecting elements are always active, and the RC vectors of different TSs are orthogonal to each other. In \cite{an_jensen_2020}, the authors provided a method to design the RC by the DFT matrix, which not only satisfies the established constraints, but also reaches the CRLB of the channel estimates. As stated in \cite{an_jensen_2020}, the minimum number of pilots required for estimating $N$ reflected BS-RIS-user channels plus a direct BS-user channel and the MSE of channel estimates are respectively given by
\begin{equation}\label{eq5-15}
\tau \left| {_{DFT}} \right. = N + 1,\quad \sigma _d^2\left| {_{DFT}} \right. = \frac{{{\sigma ^2}}}{{N + 1}},\quad \sigma _{r,n}^2\left| {_{DFT}} \right. = \frac{{{\sigma ^2}}}{{N + 1}},n = 1,2, \cdots ,N.
\end{equation}
As a result, the corresponding ergodic achievable rate for these three channel estimation schemes can be obtained by substituting (\ref{eq5-13}) $\sim$ (\ref{eq5-15}) into (\ref{eq5-12}), which is omitted here for brevity.

In summary, although several channel estimation algorithms attempted to reduce the pilot overhead, they tend to exhibit a poorer channel estimation accuracy, as well as achievable rate than our method, especially in slowing time-varying channels having longer coherence time. However, in the context of rapidly time-varying channels of UAV-assisted or vehicle-to-everything (V2X) systems, it is particularly desirable to reduce the pilot overhead, because it is doubled whenever the Doppler frequency is doubled. On the other hand, from an information-theoretic perspective, having continuously open reflecting elements can provide more observation information about the direct BS-user channel, which generally translates into improved estimation accuracy. In that sense, the ON/OFF strategy of the RIS elements not only fails to take advantage of this benefit, but also has to rely on inaccurate direct BS-user link estimates for estimating the reflected BS-RIS-user channels, which precipitates error propagation. Nevertheless, the specific characteristics of reflected BS-RIS-user channels have been exploited by these existing channel estimation strategies. For example, the sparsity of the BS-RIS channel \cite{2019arXiv191203619C}, the correlation of the reflected BS-RIS-user channels \cite{you2020channel} and the similarity of reflected BS-RIS-user channels of multiple users \cite{wang2020channel}, have been capitalized on, but the direct BS-user channel either has been ignored or a crude ON/OFF method has been adopted. In summary, although the recent channel estimation methods have succeeded in reducing the pilot overhead, it is still an open problem, as to whether techniques may be found, which can rely on always-on reflecting elements. This question is set aside for our future research.

\subsection{The Theoretical Analysis of Our Channel Training Protocol}\label{s5-2}
Next, we consider the effect of channel estimation errors on our channel training protocol. First, we assume that the channel estimate in the $q$th training period is
\begin{equation}\label{eq5-16}
{\hat h_q} = {h_q} + {\varepsilon _q},\quad q = 1,2, \cdots ,Q,
\end{equation}
where ${\varepsilon _q} \sim {\mathcal{CN}}\left( {0,{\sigma_Q ^2}} \right)$ denotes the estimation error of the superimposed channel in the $q$th training period.

Therefore, the average received power of our channel training protocol can be expressed as
\begin{equation}\label{eq5-17}
        {P_u} = P{\mathbb{E}}\left\{ {{{\left| {{\varphi _{\hat q}}{h_r} + {h_d}} \right|}^2}} \right\}= P{\mathbb{E}}\left\{ {{{\left| {{h_r}} \right|}^2} + {{\left| {{h_d}} \right|}^2} + 2\Re \left( {\varphi _{\hat q}^ * h_r^ * {h_d}} \right)} \right\},
\end{equation}
where
\begin{equation}\label{eq5-18}
    \hat q = \arg \mathop {\max }\limits_{q = 1,2, \cdots ,Q} \left\{ {{{\left| {{\varphi _q}{h_r} + {h_d} + {\varepsilon _q}} \right|}^2}} \right\},
\end{equation}
represents the optimal training period index selected from $Q$ candidates based on the estimated superimposed channel.

We note that evaluating the third term of (\ref{eq5-17}) is not trivial even for a RIS having only a single reflecting element. Hence, in this paper, we derive an upper bound for characterizing our channel training protocol in the presence of noise. For the sake of brevity, we only consider our channel training protocol relying on random configuration in this part. More specifically, the third term of (\ref{eq5-17}) is upper bounded by
\begin{equation}\label{eq5-19}
    \begin{split}
&{\mathbb{E}}\left( {\Re \left( {\varphi _{\hat q}^ * h_r^ * {h_d}} \right)\left| {{\varphi _{\hat q}} = \arg \mathop {\max }\limits_{{\varphi _q},q = 1,2, \cdots ,Q} \left\{ {{{\left| {{\varphi _q}{h_r} + {h_d} + {\varepsilon _q}} \right|}^2}} \right\}} \right.} \right)\\
 &\le {\mathbb{E}}\left( {\mathop {\max }\limits_{q = 1,2, \cdots ,Q} \left\{ {\Re \left( {{{\left( {\frac{{{\varphi _q^ *}{h_r^ *} + {\varepsilon _q^ *}}}{{\left| {{\varphi _q^ *}{h_r^ *} + {\varepsilon _q^ *}} \right|}}{\varphi _q}{h_r}} \right)}^ * }{h_d}} \right)} \right\}} \right)\\
 &\le {\mathbb{E}}\left( {\left| {\frac{{h_r^ *  + {{\varepsilon _q^ * } \mathord{\left/
 {\vphantom {{\varepsilon _q^ * } {\varphi _q^ * }}} \right.
 \kern-\nulldelimiterspace} {\varphi _q^ * }}}}{{\left| {h_r^ *  + {{\varepsilon _r^ * } \mathord{\left/
 {\vphantom {{\varepsilon _r^ * } {\varphi _r^ * }}} \right.
 \kern-\nulldelimiterspace} {\varphi _r^ * }}} \right|}}{h_r}} \right|} \right){\mathbb{E}}\left( {\left| {{h_d}} \right|} \right){\mathbb{E}}\left( {\mathop {\max }\limits_{q = 1,2, \cdots ,Q} \left\{ {\cos {\theta' _q}} \right\}} \right)\overset{(b)}{=} \frac{\pi }{4}{\rho _d}\frac{{\rho _r^2}}{{\sqrt {\rho _r^2 + {\sigma ^2}} }}g\left( Q \right),
    \end{split}
\end{equation}
where we have
\begin{equation}\label{eq5-19-1}
    \cos \theta {'_q} = {{\Re \left( {{{\left( {\frac{{h_r^ *  + {{\varepsilon _q^ * } \mathord{\left/
 {\vphantom {{\varepsilon _q^ * } {\varphi _q^ * }}} \right.
 \kern-\nulldelimiterspace} {\varphi _q^ * }}}}{{\left| {h_r^ *  + {{\varepsilon _r^ * } \mathord{\left/
 {\vphantom {{\varepsilon _r^ * } {\varphi _r^ * }}} \right.
 \kern-\nulldelimiterspace} {\varphi _r^ * }}} \right|}}{\varphi _q}{h_r}} \right)}^ * }{h_d}} \right)} \mathord{\left/
 {\vphantom {{\Re \left( {{{\left( {\frac{{h_r^ *  + {{\varepsilon _q^ * } \mathord{\left/
 {\vphantom {{\varepsilon _q^ * } {\varphi _q^ * }}} \right.
 \kern-\nulldelimiterspace} {\varphi _q^ * }}}}{{\left| {h_r^ *  + {{\varepsilon _r^ * } \mathord{\left/
 {\vphantom {{\varepsilon _r^ * } {\varphi _r^ * }}} \right.
 \kern-\nulldelimiterspace} {\varphi _r^ * }}} \right|}}{\varphi _q}{h_r}} \right)}^ * }{h_d}} \right)} {\left| {{{\left( {\frac{{h_r^ *  + {{\varepsilon _q^ * } \mathord{\left/
 {\vphantom {{\varepsilon _q^ * } {\varphi _q^ * }}} \right.
 \kern-\nulldelimiterspace} {\varphi _q^ * }}}}{{\left| {h_r^ *  + {{\varepsilon _r^ * } \mathord{\left/
 {\vphantom {{\varepsilon _r^ * } {\varphi _r^ * }}} \right.
 \kern-\nulldelimiterspace} {\varphi _r^ * }}} \right|}}{\varphi _q}{h_r}} \right)}^ * }{h_d}} \right|}}} \right.
 \kern-\nulldelimiterspace} {\left| {{{\left( {\frac{{h_r^ *  + {{\varepsilon _q^ * } \mathord{\left/
 {\vphantom {{\varepsilon _q^ * } {\varphi _q^ * }}} \right.
 \kern-\nulldelimiterspace} {\varphi _q^ * }}}}{{\left| {h_r^ *  + {{\varepsilon _r^ * } \mathord{\left/
 {\vphantom {{\varepsilon _r^ * } {\varphi _r^ * }}} \right.
 \kern-\nulldelimiterspace} {\varphi _r^ * }}} \right|}}{\varphi _q}{h_r}} \right)}^ * }{h_d}} \right|}},
\end{equation}
following $\theta {'_q} \sim U\left[ {0,2\pi } \right)$ and ${{{\varepsilon _q}} \mathord{\left/
 {\vphantom {{{\varepsilon _q}} {{\varphi _q}}}} \right.
 \kern-\nulldelimiterspace} {{\varphi _q}}} \sim {\mathcal{CN}}\left( {0,\sigma _Q^2} \right)$, {\color{black}{while (b) holds as the results of Appendix C.}}

Furthermore, analogously to the proof of \textbf{\emph{Proposition 3}}, we directly summarize the average received power of our channel training protocol in the face of noise in \textbf{\emph{Proposition 5}}.

\textbf{\emph{Proposition 5:}} Upon assuming ${h_{r,n}} \sim {\cal C}{\cal N}\left( {0,\rho _r^2} \right),n = 1,2, \cdots ,N$, ${h_d} \sim {\mathcal{CN}}\left( {0,\rho _d^2} \right)$ and ${\varepsilon _q} \sim {\mathcal{CN}}\left( {0,\sigma _Q^2} \right)$, we have:
\begin{equation}\label{eq5-20}
    {P_u} \le P\left( {N\rho _r^2 + \rho _d^2 + \frac{\pi }{2}N\frac{{\rho _r^2}}{{\sqrt {\rho _r^2 + {\sigma_Q ^2}} }}{\rho _d}g\left( Q \right) + \frac{\pi }{4}N\left( {N - 1} \right)\frac{{\rho _r^4}}{{\rho _r^2 + {\sigma_Q ^2}}}{g^2}\left( Q \right)} \right).
\end{equation}

By comparing \textbf{\emph{Proposition 4}} and \textbf{\emph{Proposition 5}}, one can see that although our channel training protocol suffers from some loss of power, it is more robust with respect to the estimation errors than the channel estimation protocol, which is consistent with our intuition. More specifically, in the channel estimation protocol of \cite{you2020channel, an_jensen_2020, wang2020channel, channel_mishra_2019, 2019arXiv191203619C, zhang2019capacity, guo2019weighted, cui2019secure, you2020intelligent, zheng2019intelligent}, both the estimated direct BS-user channel and the estimated reflected BS-RIS-user channel are inaccurate, which has a severe impact on the optimization of the RIS RC. By contrast, in our channel training protocol, the estimation error has a reduced influence on the overall performance. On the other hand, although the average received power of our channel training protocol is upper-bounded, the following simulations will show that \textbf{\emph{Proposition 5}} represents a rather tight upper bound at low SNRs and confirms the robustness of our channel training protocol.
%%%%%%%%%%%%%%%%%%%%%%%%%%%%%%%%%%%%%%%%%%%%%%%%%%%%%%%%%%%%% Simulation
\section{Simulation Results}\label{s6}
\begin{figure}[!t]
	\centering
	\includegraphics[width=7cm]{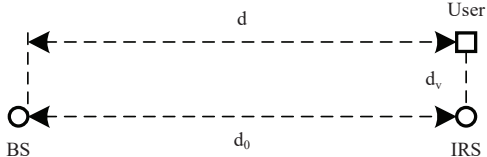}
	\caption{The position schematic of the single-user scenario (top view).}
	\label{f5}
\end{figure}

In this section, numerical simulations are provided for characterizing the proposed protocols. In our simulations, we consider a three-dimensional (3D) coordinate system where a uniform linear array (ULA) is used at the BS and a uniform rectangular array (URA) is employed at the RIS, which are located in the $x$-axis and $x - z$ plane, respectively. The antenna spacing is half a wavelength and the center of the antennas at the BS and RIS is located at $\left( {0,0,0} \right)$ and $\left( {0,{d_0},0} \right)$, respectively, where ${d_0}$ is set to ${d_0} = 50$ meter (m) denoting the distance between the BS and the RIS. For the number of RIS elements, we set $N = {N_x}{N_z}$, where ${N_x}$ and ${N_z}$ denote the numbers of reflecting elements along the $x$-axis and $z$-axis, respectively, where we fix ${N_x} = 10$. Furthermore, we assume a Rician fading channel model for all channels involved. More specifically, the BS-RIS channel ${\bf{U}}$ is given by \cite{wu2019intelligent}
 \begin{equation}\label{eq6-1}
{\bf{U}} = \sqrt {{C_0}{{d_{BI}}^{ - {\alpha_{BI}} }}} \left( {\sqrt {\frac{{{\beta _{BI}}}}{{{\beta _{BI}} + 1}}} {{\bf{U}}_{LoS}} + \sqrt {\frac{1}{{{\beta _{BI}} + 1}}} {{\bf{U}}_{NLoS}}} \right),
 \end{equation}
where ${C_0}$ is set to ${C_0} =  - 20$ dB, denoting the path loss at the reference distance of $1$ m, $d_{BI}=d_0$ denotes the link length between the BS and RIS, and $\alpha_{BI} $ denotes the corresponding path loss exponent, ${{\beta _{BI}}}$ is the Rician factor, and ${{\bf{U}}_{LoS}}$ and ${{\bf{U}}_{NLoS}}$ represent the deterministic LoS and Rayleigh fading components, respectively. In particular, the above model is reduced to the LoS channel when ${\beta _{BI}} \to \infty $ or to the Rayleigh fading channel, when ${\beta _{BI}} = 0$. The BS-user and RIS-user channels are also generated by following a similar procedure to that of (\ref{eq6-1}). The path loss exponents of the BS-user and RIS-user links are denoted by ${\alpha _{BU}}$ and ${\alpha _{IU}}$, respectively, and the Rician factors of the two links are denoted by ${\beta _{BU}}$ and ${\beta _{IU}}$, respectively. Due to the relatively large distance and random scattering of the BS-user channel, we set ${\alpha _{BU}} = 3.5$ and ${\beta _{BU}} = 0$, unless specified otherwise, while their counterparts for the BS-RIS and RIS-user channels will be specified later to study their effects on the system performance. In our simulations, we assume that all users have the same SINR target, i.e., ${\gamma _k} = \gamma ,k = 1,2, \cdots ,K$. The number of random vectors used for the Gaussian randomization in the SDR algorithm is set to 1000 and the stopping threshold for the alternating optimization algorithms is set to $\varepsilon  = {10^{ - 4}}$. The noise power at the receiver is set as $\sigma _n^2 =  - 70$ dBm for $k = 1,2, \cdots ,K$.

\begin{figure}[!t]
	\centering
	\includegraphics[width=7cm]{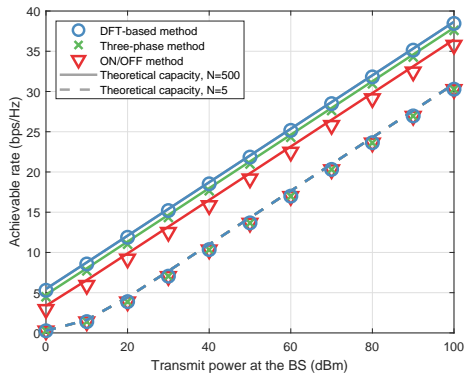}
	\caption{The achievable rate versus the transmit power at the BS for the different channel estimation methods of Section \ref{s3}, where the average power of pilot symbols is $-30$ dBm.}
	\label{f6}
\end{figure}
\begin{figure}[!t]
	\centering
	\includegraphics[width=7cm]{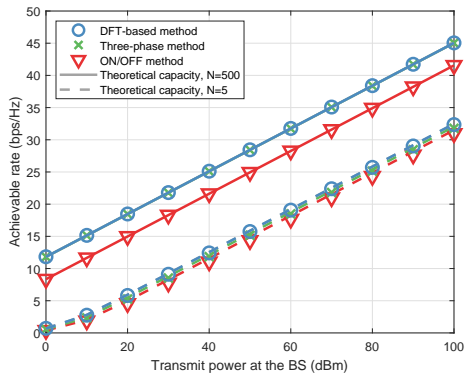}
	\caption{The achievable rate versus the transmit power at the BS for the different channel estimation methods of Section \ref{s3}, where the average power of pilot symbols is $10$ dBm.}
	\label{f7}
\end{figure}
\subsection{SISO Scenario}\label{s6-1}
First, let us consider the SISO scenario for verifying the theoretical analysis in this paper, where a single user is served by the BS with a single TA. As shown in Fig. \ref{f5}, the user lies on a horizontal line that is in parallel to the BS-RIS link, where the vertical distance between these two lines is set to ${d_v} = 3$ m, and the horizontal length between the BS and the user is denoted by $d$ m. Accordingly, the lengths of the BS-user and RIS-user links are given by ${d_{BU}} = \sqrt {{d^2} + d_v^2} $ and ${d_{IU}} = \sqrt {{{\left( {{d_0} - d} \right)}^2} + d_v^2} $, respectively. The path loss exponents and Rician factors are set to ${\alpha _{BI}} = 2$, ${\alpha _{IU}} = 2.8$, ${\beta _{BI}} = \infty $, and ${\beta _{IU}} = 0$, respectively, which results in ${\bf{U}}$ having a rank of one, i.e. an LoS channel between the BS and the RIS. The noise power at the BS is set to $\sigma _z^2 =  - 70$ dBm.
\subsubsection{Achievable rate of the channel estimation protocol considering noise at the BS} In Figs. \ref{f6} $\sim$ \ref{f7}, we first study the achievable rate of the channel estimation protocol considering realistic channel estimation errors at the BS, where the three different channel estimation methods described in Section \ref{s3} are considered. In Fig. \ref{f6}, the average power of the pilot symbols used for estimating the channels is set to $\alpha =  - 30$ dBm. Observe from Fig. \ref{f6} that the channel estimation methods have a significant influence on the achievable DL rate. The DFT-based channel estimation method has the most accurate estimator, and accordingly its achievable rate is also the highest. The three-phase and ON/OFF methods exhibit a power loss of $1.5$ dB and $8.5$ dB, respectively, when $N = 500$ reflecting elements are installed at the RIS, which is consistent with our theoretical analysis. Nevertheless, for a small number of RIS elements, the difference in the achievable rates between the three estimators is not significant, because only a limited performance improvement is achieved by the RIS. As the number of reflecting elements increases, the rate loss becomes more pronounced. For a large number of RIS elements, the ON/OFF method has a severe estimation error, which results in a rate loss compared to the optimal DFT-based estimation method.

In Fig. \ref{f7}, we increase the power of the pilot symbols used for estimating the channels to $\alpha =  10$ dBm, which results in more accurate channel estimates. It can be seen from Fig. \ref{f7} that the three-phase method behaves almost as well as the DFT-based method, which also verifies our Corollary 4. Explicitly, the achievable rate, i.e., the average received power, is determined by the estimation accuracy of the reflected BS-RIS-user channels when a large number of reflecting elements is installed. This also explains why the ON/OFF method still suffers from a rate loss of around $8$ dB. Additionally, we can observe by comparing Figs. \ref{f6} and \ref{f7} that with the increase of the pilot symbol power, the achievable DL rate is also improved accordingly. Therefore, how to reduce rate loss by balancing the power between the pilot symbols and the information symbols is crucial for the channel estimation protocol, especially for a large number of reflecting elements.
\subsubsection{Average received power of our channel training protocol without noise at the BS}

\begin{figure}[!t]
	\centering
	\includegraphics[width=7cm]{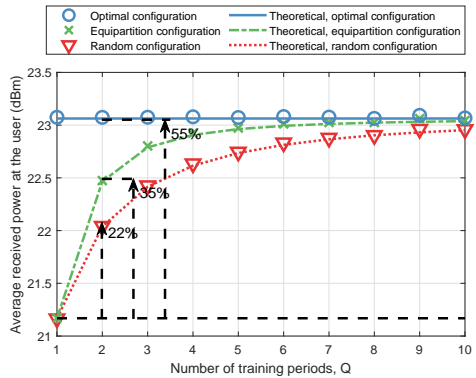}
	\caption{The average received power of our channel training protocol versus the number of training periods in the absence of noise, where the number of reflecting elements is set to $N=1$.}
	\label{f8}
\end{figure}
\begin{figure}[!t]
	\centering
	\includegraphics[width=7cm]{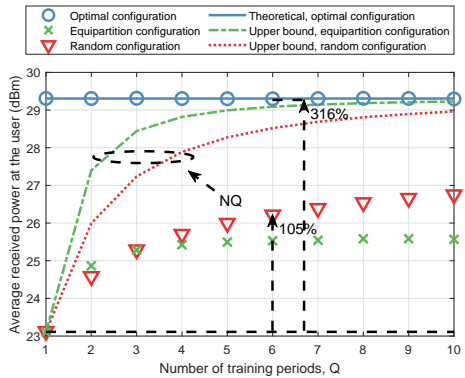}
	\caption{The average received power of our channel training protocol versus the number of training periods in the absence of noise, where the number of reflecting elements is set to $N=5$.}
	\label{f9}
\end{figure}
Figs. \ref{f8} and \ref{f9} verify the efficiency of our channel training protocol for the SISO scenario in terms of the average received power versus the number of training periods. In Fig. \ref{f8}, we consider only a single reflecting element in order to verify \textbf{\emph{Proposition 2}}. As seen from Fig. \ref{f8}, upon increasing the number of training periods, the performance of our channel training protocol will gradually approach the performance of the optimal RIS configuration. On the other hand, the equi-partition configuration has a faster convergence than the random configuration method. More specifically, let us consider the fair comparison condition of $Q = 2$, which is said to be fair, because at least two pilot symbols are required for completing the CSI acquisition for the channel estimation protocol of \cite{you2020channel, an_jensen_2020, wang2020channel, channel_mishra_2019, 2019arXiv191203619C, zhang2019capacity, guo2019weighted, cui2019secure, you2020intelligent, zheng2019intelligent}. In particular, $Q = 1$ represents the case of a single random configuration. It can be seen from Fig. \ref{f8} that, compared to the single random configuration, channel training relying on the random configuration and on the equi-partition configuration is capable of increasing the average received power by 22\% and 35\%, respectively, while the optimal configuration increase the average received power by 55\%. Although our channel training scheme has a substantial performance improvement over the single random configuration, it still has a certain amount of power loss compared to the optimal configuration. However, it should be noted that Fig. \ref{f8} does not take into account the impact of channel estimation errors, which have a more grave impact on the channel estimation protocol of \cite{you2020channel, an_jensen_2020, wang2020channel, channel_mishra_2019, 2019arXiv191203619C, zhang2019capacity, guo2019weighted, cui2019secure, you2020intelligent, zheng2019intelligent}. In the later simulations, we will see the competitiveness of our channel training protocol in a noisy wireless environment.

In Fig. \ref{f9}, we consider the general case of having $N$ number of reflecting elements, where we set $N = 5$. Observe from Fig. \ref{f9} that the situation is slightly different from that in Fig. \ref{f8}. First of all, compared to the theoretical derivation, the training protocol suffers from a severe power loss, which is caused by the scaling in (\ref{eq4-23}). Furthermore, the equi-partition configuration also exhibits a significant performance loss. This is because compared to the random configuration, the equi-partition configuration has a degree of freedom of $1$. Specifically, if the initial phase shift vector is not a good choice, the training protocol using the equi-partition configuration fails to improve the performance effectively, while the random configuration has the opportunity to avoid this suboptimality in the following $\left( {Q - 1} \right)$ training periods\footnote{Although the simulation results demonstrate that the performance improvement of the equi-partition configuration scheme is only non-negligible for a small value of $N$, it is worth noting that we are considering the case of independent reflected BS-RIS-user channels. When reflected BS-RIS-user channels associated with high correlation, e.g., ultra-dense reflecting elements, are considered, the equi-partition configuration still works effectively.}. Following the same comparison principle in Fig. \ref{f8}, the random configuration increases the average received power by $105$\%, but fails to approach the $316$\% improvement of the optimal configuration. On the other hand, the upper bound is achievable if we do not consider the limitations of the training periods. For example, considering the configuration of only a single reflecting element at a given time in all $Q$ training periods, with all other reflecting elements turned off, the upper bound becomes achievable, when doing so for $N$ elements. Therefore, reaching the upper bound requires at most $NQ$ training periods. This also verifies that the upper bound is tight for $N = 1$, as shown in \textbf{\emph{Proposition 1}} and Fig. \ref{f8}. Nevertheless, how to design a good training method requiring a limited number of training periods for effectively improving the performance of the RIS is still an open subject, which will be set aside for our future research.

 \begin{figure}[!t]
	\centering
	\includegraphics[width=7cm]{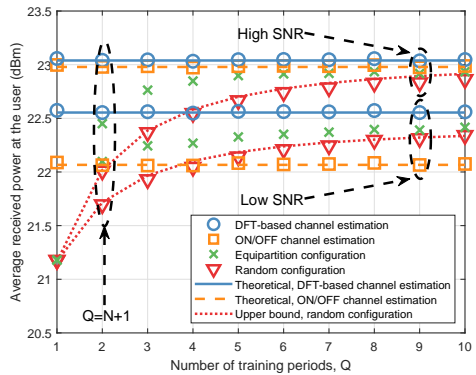}
	\caption{The average received power of our channel training protocol versus the number of training periods in the face of noise, where the number of reflecting elements is set to $N=1$.}
	\label{f10}
\end{figure}
\begin{figure}[!t]
	\centering
	\includegraphics[width=7cm]{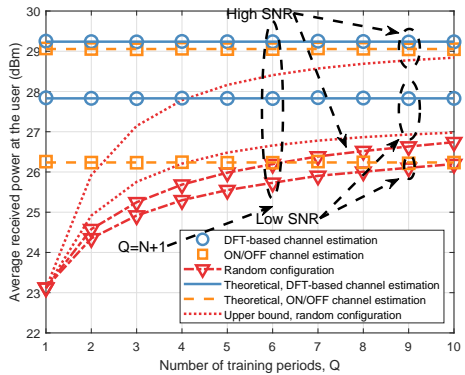}
	\caption{The average received power of our channel training protocol versus the number of training periods in the face of noise, where the number of reflecting elements is set to be $N=5$.}
	\label{f11}
\end{figure}
\subsubsection{Average received power of our channel training protocol in the presence of noise at the BS} Figs. \ref{f10} and \ref{f11} compare the performance of the channel estimation protocol of \cite{you2020channel, an_jensen_2020, wang2020channel, channel_mishra_2019, 2019arXiv191203619C, zhang2019capacity, guo2019weighted, cui2019secure, you2020intelligent, zheng2019intelligent} and our channel training protocol in the face of noise. Similarly, the average received power is adopted as the evaluation criterion in order to compare our theoretical analysis and the DFT-based method as well as the ON/OFF method. The average power of the pilot symbols used for estimating the channel is set to $\alpha = 15$ dBm and $\alpha = 30$ dBm, respectively, corresponding to the labels of low SNR and high SNR in Figs. \ref{f10} and \ref{f11}. It can be seen from Fig. \ref{f10} that all the performance curves have similar trends to those in Fig. \ref{f8}. Upon increasing the number of training periods, the performance of our channel training protocol will be improved. Additionally, the performance of the ON/OFF channel estimation method suffers from a higher power loss than that of the DFT-based method. In particular, it is worth noting that our channel training protocol is most competitive under low SNRs. More specifically, considering the same comparison strategy as described for Fig. \ref{f8}, when $Q = 2$ and low SNR are considered, our channel training based on the equi-partition configuration outperforms the channel estimation protocol of \cite{channel_mishra_2019} based on the ON/OFF method. Even compared to the DFT-based channel estimation method of \cite{an_jensen_2020}, our equi-partition based channel training protocol suffers from a smaller power loss than that shown in Fig. \ref{f8}.

\begin{figure}[!t]
	\centering
	\includegraphics[width=7cm]{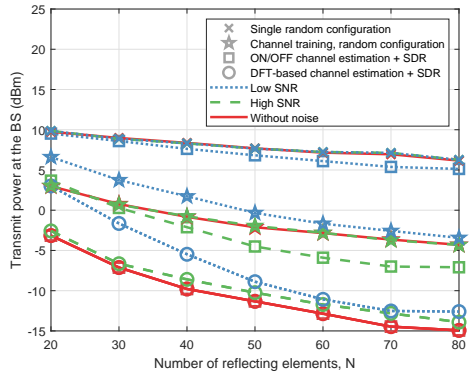}
	\caption{The transmit power at the BS versus the number of reflecting elements, where we have $d=50$ m.}
	\label{f12}
\end{figure}
\begin{figure}[!t]
	\centering
	\includegraphics[width=7cm]{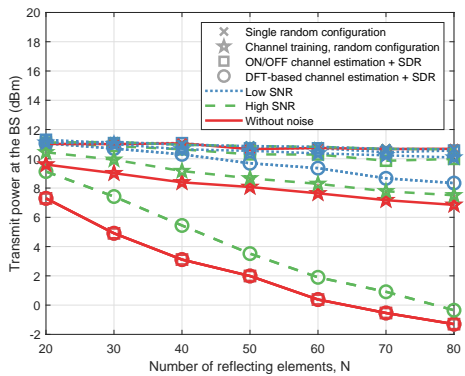}
	\caption{The transmit power at the BS versus the number of reflecting elements, where we have $d=40$ m.}
	\label{f13}
\end{figure}
In Fig. \ref{f11}, we increase the number of reflecting elements to $N = 5$ and continue to study the effects of noise, while other simulation conditions are consistent with Fig. \ref{f10}. Due to the deficiency of the equi-partition configuration in the case of a large number of reflecting elements, we only consider the random configuration in Fig. \ref{f11}. The theoretical upper bound of our channel training protocol in Fig. \ref{f11} comes from our derivation in Section \ref{s5}. One can observe from Fig. \ref{f11} that, in the low-SNR region, our channel training protocol is quite competitive compared to its ON/OFF channel estimation counterpart. In addition, an implicit phenomenon can be observed in Fig. \ref{f11}, namely that compared to the channel estimation protocol, our channel training protocol is more robust to the estimation error, which is consistent with our previous theoretical analysis.
%%%%%%%%%%%%%%%%%%%%%%%%%%%%%%%%%%%%%%%%%%%%%%%%%%%%%%%%%%%%%%%%%%%%%%%%%%%%%%%% MISO scenario
\subsection{MISO Scenario}\label{s6-2}
\begin{figure}[!t]
	\centering
	\includegraphics[width=7cm]{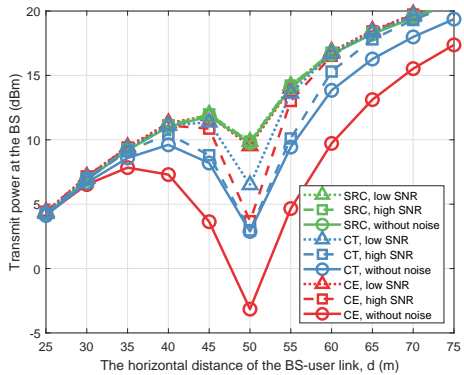}
	\caption{The transmit power at the BS versus the horizontal length of the BS-user link.}
	\label{f14}
\end{figure}
\subsubsection{Transmit power versus the number of reflecting elements}
Next, let us consider the case where the BS is equipped with $M = 4$ TAs. The relative positions of the BS, RIS and the user is still shown in Fig. \ref{f5}. The target SINR is set to $\gamma  = 10$ dB. In Figs. \ref{f12} and \ref{f13}, we compare the transmit power of our channel training protocol and the channel estimation protocol of \cite{you2020channel, an_jensen_2020, wang2020channel, channel_mishra_2019, 2019arXiv191203619C, zhang2019capacity, guo2019weighted, cui2019secure, you2020intelligent, zheng2019intelligent}. For the channel estimation protocol, the ON/OFF estimation method and the DFT-based estimation method are adopted, and the SDR algorithm is used for optimizing the RCs \cite{wu2019intelligent}. For our channel training protocol, we consider the random configuration training method. For the sake of fairness, we set the number of training periods as $Q = N + 1$, which is the minimum number of pilot symbols required for the CSI acquisition in the channel estimation protocol \cite{you2020channel, an_jensen_2020, wang2020channel}. In addition, the performance of the single random configuration is plotted both in Fig. \ref{f12} and \ref{f13} for comparison. In Fig. \ref{f12}, we consider $d = 50$ m, which corresponds to the case where the user is closest to the RIS. Observe from Fig. \ref{f12} that, upon increasing the number of reflecting elements, the BS achieves the expected SINR target at a reduced transmit power. In particular, the relationship of the square law under the optimal configuration is also shown in Fig. \ref{f12}. For example, without considering the noise at the BS, the transmit power decreases from $ - 7$ dBm to $ - 13$ dBm when increasing the number of RIS elements from $N = 30$ to $N = 60$, gaining about 6dB. Fig. \ref{f12} demonstrate that in the ideal case without noise, our channel training protocol suffers about 7dB loss compared to the channel estimation protocol, when we have $N = 30$. However, the situation is slightly different in the presence of noise. When considering realistic estimation errors, the performance of all protocols will be eroded. Comparatively speaking, our channel training protocol is most robust to estimation errors. This is because even if the index of the selected objective value at the decision centre is not optimal, it is still near that of the optimal value. More specifically, our channel training protocol outperforms the ON/OFF channel estimation method at low SNRs. Even compared to the DFT-based estimation method, our channel training protocol only suffers from a power penalty of about 3dB at $N = 20$, despite it's reduced system complexity and pilot overhead.

In Fig. \ref{f13}, we consider the case of $d = 40$ m, while all other simulation setups are the same as in Fig. \ref{f12}. Upon comparing Figs. \ref{f12} and \ref{f13}, it can be seen that as the user moves away from the RIS, more transmit power is required to achieve the same SNR target. Additionally, observe from Fig. \ref{f13} that even the most robust DFT-based estimation method will suffer severe power loss at low SNRs, because as the user moves away from the RIS, more accurate channel estimates are required to facilitate coherent superposition of the direct and reflected signals. At low SNRs, the performance improvement of the RIS has degraded significantly, and our channel training protocol performs almost as well as the channel estimation protocol of \cite{you2020channel, an_jensen_2020, wang2020channel, channel_mishra_2019, 2019arXiv191203619C, zhang2019capacity, guo2019weighted, cui2019secure, you2020intelligent, zheng2019intelligent}.
\subsubsection{Transmit power versus RIS-User distance}
In Fig. \ref{f14}, we compare of transmit power required by different protocols versus the BS-user link length, $d$, in the face of noise. For the sake of illustration, we consider the ON/OFF channel estimation method in Fig. \ref{f14}. One can see that, in the vicinity of the RIS, there is a significant reduction of the transmit power. Furthermore, it can be observed from Fig. \ref{f14} that, in the ideal case without noise, our channel training protocol suffers from some performance loss compared to channel estimation protocol of \cite{you2020channel, an_jensen_2020, wang2020channel, channel_mishra_2019, 2019arXiv191203619C, zhang2019capacity, guo2019weighted, cui2019secure, you2020intelligent, zheng2019intelligent}. For example, for the same transmit power of $9$ dBm, the coverage range of the channel estimation protocol is about $59$ m, while the range of our channel training protocol is reduced to $55$ m. Nevertheless, in the presence of estimation errors the situation is quite different. Our channel training protocol outperforms the channel estimation protocol at the same level of errors. For example, in the case of $d=50$ m, the transmit power of 7dBm is adequate when considering low SNRs, while 10dB is required by the channel estimation protocol, which once again confirms the robustness of our channel training protocol under estimation errors, especially at low SNRs.

%%%%%%%%%%%%%%%%%%%%%%%%%%%%%%%%%%%%%%%%%%%%%%%%%%%%%%%%%%%%%%%%%%%%%%%%% MIMO scenario
\subsection{MIMO Scenario}\label{s6-3}
\begin{figure}[!t]
	\centering
	\includegraphics[width=7cm]{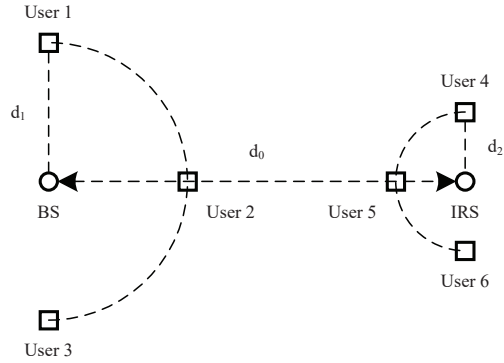}
	\caption{The position schematic of the multiuser scenario (top view).}
	\label{f15}
\end{figure}

Next, let us consider a multi-user system supporting six users. The positions of users relative to the BS and RIS are shown in Fig. \ref{f15}, where users 1, 2, and 3 lie on a semicircle with BS as the centre, and the radius is set to $d = 15$ m. Users 4, 5, 6 lie on a semicircle centred on the RIS with radius of ${d_2} = 3$ m. Since user 1, 2, 3 are far away from the RIS, we set ${\alpha _{IU}} = 3.5$ for users $1 \sim 3$ and ${\alpha _{IU}} = 2.8$ for users $4 \sim 6$, respectively. Following the same consideration, we set ${\alpha _{BU}} = 2.8$ for users $1 \sim 3$ and ${\alpha _{BU}} = 3.5$ for users $4 \sim 6$, respectively. Since the RIS-BS channel is a LoS link, we set ${\beta _{BI}} = 5$ dB. Both the three-phase method and the DFT-based method are considered. For the three-phase method, we consider the user 1 as the specific user. The problem (\ref{eq3-6}) of joint multi-user TPC and RC is solved by the alternate optimization algorithm of \cite{wu2019intelligent}. More specifically, the multi-user TPC problem is solved by using the fixed-point iterative algorithms of \cite{linear_wiesel_2006}, while the optimization of the RC is solved by using the SDR algorithm of \cite{wu2019intelligent}. In \cite{wu2019intelligent}, the convergence of alternating optimization algorithms has been theoretically analyzed and verified. Similarly, in our channel training protocol, we use the algorithm of \cite{linear_wiesel_2006} for solving the multi-user TPC during each training period. The average pilot power is set to $15$ dBm and $30$ dBm, corresponding to the case of low and high SNRs, respectively.
\subsubsection{Transmit power versus the number of reflecting elements}
\begin{figure}[!t]
	\centering
	\includegraphics[width=7cm]{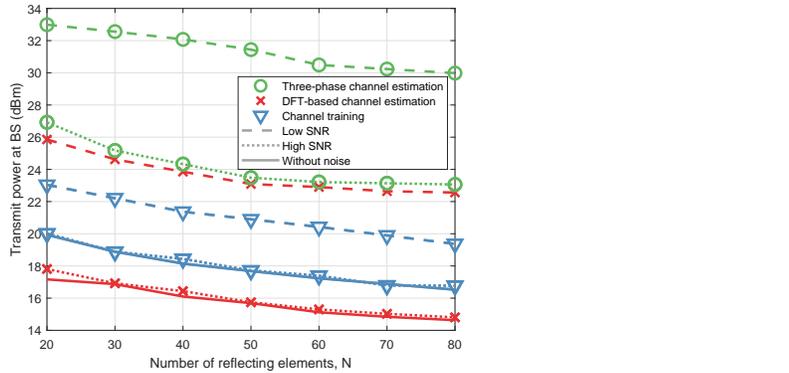}
	\caption{The transmit power at the BS versus the number of reflecting elements for $K=6$ users.}
	\label{f16}
\end{figure}

First, the transmit power versus the number of RIS elements is shown in Fig. \ref{f16}, where the gain of the RIS is much lower than that of the single-user scenario. For example, the power gain is only about 3dB upon increasing the number of RIS elements from $N = 20$ to $N = 80$, compared to the theoretical 12dB observed from Fig. \ref{f16} for a single user. This is because the RIS only has to align the reflected BS-RIS-user channel with the direct BS-user channel in a single-user scenario, while in a multi-user scenario, the RIS additionally has to eliminate the multi-user interference. On the other hand, the users that are far away from the RIS are less likely to benefit from it. Therefore, more distributed RISs should be used for serving multiple users in a cell. More significantly, our channel training protocol outperforms the channel estimation protocol of \cite{you2020channel, an_jensen_2020, wang2020channel, channel_mishra_2019, 2019arXiv191203619C, zhang2019capacity, guo2019weighted, cui2019secure, you2020intelligent, zheng2019intelligent}, when the estimation errors are severe for multiple users. For example, when considering the average pilot power of 15 dBm, 3dB power gain is attained by our channel training protocol over the DFT-based channel estimation protocol. Additionally, although the three-phase method has the advantage of reducing the pilot overhead, it suffers from a severe performance loss compared to both the optimal RCs and to our channel training protocol due to the severe estimation error propagation between the channel estimates of different users.

\subsubsection{Transmit power versus user SINR target}
\begin{figure}[!t]
	\centering
	\includegraphics[width=7cm]{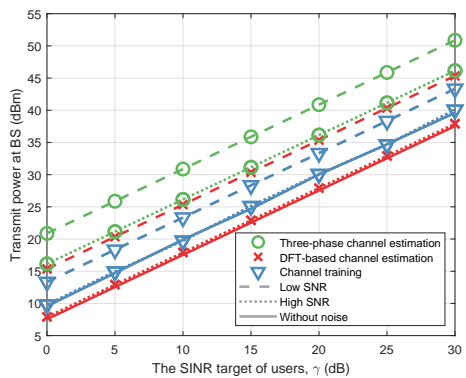}
	\caption{The transmit power at the BS versus the SINR target of $K=6$ users.}
	\label{f17}
\end{figure}

Fig. \ref{f17} shows the variation of transmit power at the BS versus the target SINR of users. The number of RIS reflecting element is set to $N = 20$. Observe that, as the SINR target increases, more transmit power is required at the BS. Additionally, the same conclusions can be drawn by comparing different protocols. For example, our channel training protocol has advantages over all channel estimation algorithms at low SNRs. Given the improvement of the channel estimates, the channel estimation protocol of \cite{you2020channel, an_jensen_2020, wang2020channel, channel_mishra_2019, 2019arXiv191203619C, zhang2019capacity, guo2019weighted, cui2019secure, you2020intelligent, zheng2019intelligent} still has a performance advantage of about 2dB.

\section{Conclusions}\label{s7}
In this paper, we proposed a novel channel training protocol for RIS-assisted multi-user communications. In contrast to the existing channel estimation protocols of \cite{you2020channel, an_jensen_2020, wang2020channel, channel_mishra_2019, 2019arXiv191203619C, zhang2019capacity, guo2019weighted, cui2019secure, you2020intelligent, zheng2019intelligent}, our channel training protocol avoids the estimation of the cascaded channels and instead estimates the superimposed channels including the direct BS-user channel and the reflected BS-RIS-user channels. The phase configuration of the RCs is accomplished by comparing the objective values over multiple training periods. Moreover, we proposed a pair of configuration strategies for our channel training protocol and discussed the corresponding theoretical performance of these methods. Additionally, we also analysed the theoretical performance of the channel estimation protocols of \cite{you2020channel, an_jensen_2020, wang2020channel, channel_mishra_2019, 2019arXiv191203619C, zhang2019capacity, guo2019weighted, cui2019secure, you2020intelligent, zheng2019intelligent} and of our channel training protocol in the face of estimation errors. Finally, the simulation results verified our conclusions. Specifically, when realistic practical estimation errors are considered, our channel training protocol shows clear performance vs. implementation complexity, pilot overhead and signalling overhead advantages.

Nevertheless, there is a range of open topics that deserve further research. For example, in practical systems, the phase shift of RIS elements is of finite precision, and the performance of our channel training protocols using finite-precision RIS elements has to be verified. Additionally, it is also interesting to see, whether there are better training methods, perhaps relying on deep learning. Furthermore, please recall that we conceive a selection mechanism that gleans information from $Q$ training periods. It may be beneficial to design an information fusion center for combining the information of $Q$ training periods and thus improve the training performance.

\appendices
\section{Proof of \textbf{\emph{Proposition 1}}}\label{A1}
First, the average channel gain in (\ref{eq4-10}) can be expanded as
\begin{equation}\label{eq4-13}
    \begin{split}
\mathbb{E}\left\{ {\mathop {\max }\limits_{q = 1,2, \cdots ,Q} \left\{ {{{\left| {{\varphi _q}h_r^* + h_d^*} \right|}^2}} \right\}} \right\}&= \mathbb{E}\left\{ {\mathop {\max }\limits_{q = 1,2, \cdots ,Q} \left\{ {{{\left| {h_r^*} \right|}^2} + 2\Re \left\{ {\varphi _q^*{h_r}h_d^*} \right\} + {{\left| {h_d^*} \right|}^2}} \right\}} \right\}\\
 &= \mathbb{E}\left\{ {{{\left| {h_r^*} \right|}^2} + \mathop {\max }\limits_{q = 1,2, \cdots ,Q} \left\{ {2\left| {{h_r}} \right|\left| {{h_d^*}} \right|\cos {\theta _q}} \right\} + {{\left| {h_d^*} \right|}^2}} \right\}\\
 &\overset{(a)}{=} \mathbb{E}\left\{ {{{\left| {h_r^*} \right|}^2}} \right\} + \mathbb{E}\left\{ {{{\left| {h_d^*} \right|}^2}} \right\}+ 2\mathbb{E}\left\{ {\left| {{h_r}} \right|} \right\}\mathbb{E}\left\{ {\left| {{h_d^*}} \right|} \right\}\mathbb{E}\left\{ {\mathop {\max }\limits_{q = 1,2, \cdots ,Q} \left\{ {\cos {\theta _q}} \right\}} \right\},
\end{split}
\end{equation}
where (a) holds because ${\varphi _q}$, ${h_r}$ and ${h_d}$ are independent of each other, while $\cos {\theta _q} = \frac{{\Re \left( {\varphi _q^*{h_r}h_d^*} \right)}}{{\left| {{h_r}} \right|\cdot\left| {h_d^*} \right|}}$ denotes the cosine of the angle between ${\varphi _q^ * {h_r}}$ and ${h_d^*}$. It is straightforward to see that ${\theta _q} \sim {\mathcal{U}}\left[ {0,2\pi } \right)$.

Note that the challenge in (\ref{eq4-13}) is to evaluate the mean of the maximum cosine function's maximum when having multiple uniformly distributed phases. Without loss of generality, let us consider one of the possible cases, which is
\begin{equation}\label{eq4-14}
    \mathop {\max }\limits_{q = 1,2, \cdots ,Q} \left\{ {\cos {\theta _q}} \right\} = \cos {\theta _1} \ge \cos {\theta _2} \ge  \cdots  \ge \cos {\theta _Q} = \mathop {\min }\limits_{q = 1,2, \cdots ,Q} \left\{ {\cos {\theta _q}} \right\}.
\end{equation}

Furthermore, due to the symmetry of the cosine function with respect to the real axis, when only considering the phase interval of $\left[ {{\rm{0,}}\pi } \right]$, the condition satisfying (\ref{eq4-14}) can be simplified by
\begin{equation}\label{eq4-15}
    0 \le {\theta _1} \le {\theta _2} \le  \cdots  \le {\theta _Q} \le \pi.
\end{equation}

Therefore, $\mathbb{E}\left\{ {\mathop {\max }\limits_{q = 1,2, \cdots ,Q} \left\{ {\cos {\theta _q}} \right\}} \right\}$ in (\ref{eq4-13}) can be expanded as
\begin{equation}\label{eq4-16}
    \mathbb{E}\left\{ {\mathop {\max }\limits_{q = 1,2, \cdots ,Q} \left\{ {\cos {\theta _q}} \right\}} \right\} = \frac{{Q!}}{{{\pi ^Q}}}\int\limits_0^\pi  {\left( { \cdots \int\limits_0^{{\theta _3}} {\left( {\int\limits_0^{{\theta _2}} {\cos {\theta _1}d{\theta _1}} } \right)d{\theta _2}}  \cdots } \right)d{\theta _Q}}.
\end{equation}

Next, we will use the mathematical induction to find the general form of (\ref{eq4-16}). First of all, the integral results of (\ref{eq4-16}) for $Q = 1,2, \cdots ,6$ are listed as follows
\begin{equation}\label{eq4-17}
    \begin{split}
&Q = 1:\int\limits_0^\pi  {\cos {\theta _1}d{\theta _1}}  = 0\\
&Q = 2:\int\limits_0^\pi  {\int\limits_0^{{\theta _2}} {\cos {\theta _1}d{\theta _1}} d{\theta _2}}  = 2\\
&Q = 3:\int\limits_0^\pi  {\int\limits_0^{{\theta _3}} {\int\limits_0^{{\theta _2}} {\cos {\theta _1}d{\theta _1}} d{\theta _2}} d{\theta _3}}  = \pi \\
&Q = 4:\int\limits_0^\pi  {\int\limits_0^{{\theta _4}} {\int\limits_0^{{\theta _3}} {\int\limits_0^{{\theta _2}} {\cos {\theta _1}d{\theta _1}} d{\theta _2}} d{\theta _3}} d{\theta _4}}  = \frac{1}{2}{\pi ^2} - 2\\
&Q = 5:\int\limits_0^\pi  {\int\limits_0^{{\theta _5}} {\int\limits_0^{{\theta _4}} {\int\limits_0^{{\theta _3}} {\int\limits_0^{{\theta _2}} {\cos {\theta _1}d{\theta _1}} d{\theta _2}} d{\theta _3}} d{\theta _4}} d{\theta _5}}  = \frac{1}{{3!}}{\pi ^3} - \pi \\
&Q = 6:\int\limits_0^\pi  {\int\limits_0^{{\theta _6}} {\int\limits_0^{{\theta _5}} {\int\limits_0^{{\theta _4}} {\int\limits_0^{{\theta _3}} {\int\limits_0^{{\theta _2}} {\cos {\theta _1}d{\theta _1}} d{\theta _2}} d{\theta _3}} d{\theta _4}} d{\theta _5}} d{\theta _6}}  = \frac{1}{{4!}}{\pi ^4} - \frac{1}{2}{\pi ^2} + 2.
\end{split}
\end{equation}

Based on (\ref{eq4-17}), we can express (\ref{eq4-16}) as follows
\begin{equation}\label{eq4-18}
    \begin{split}
&\int\limits_0^\pi  {\left\{ {\sum\limits_{i = 1}^{\left\lceil {{{\left( {Q - 2} \right)} \mathord{\left/
 {\vphantom {{\left( {Q - 2} \right)} 2}} \right.
 \kern-\nulldelimiterspace} 2}} \right\rceil } {\frac{{{{\left( { - 1} \right)}^{i + 1}}\theta _Q^{Q - 1 - 2i}}}{{\left( {Q - 1 - 2i} \right)!}}}  + \cos \left( {{\theta _Q} - \frac{{\left( {Q - 1} \right)\pi }}{2}} \right)} \right\}d{\theta _Q}}  \\
 &= \sum\limits_{i = 1}^{\left\lceil {{{\left( {Q - 2} \right)} \mathord{\left/
 {\vphantom {{\left( {Q - 2} \right)} 2}} \right.
 \kern-\nulldelimiterspace} 2}} \right\rceil } {\frac{{{{\left( { - 1} \right)}^{i + 1}}{\pi ^{Q - 2i}}}}{{\left( {Q - 2i} \right)!}}}  + 2\sin \left( {\frac{{\left( {Q - 1} \right)\pi }}{2}} \right)\\
 &= \sum\limits_{i = 1}^{\left\lceil {{{\left( {Q - 1} \right)} \mathord{\left/
 {\vphantom {{\left( {Q - 1} \right)} 2}} \right.
 \kern-\nulldelimiterspace} 2}} \right\rceil } {{{\left( { - 1} \right)}^{i + 1}}f\left( {Q - 2i} \right){\pi ^{Q - 2i}}},
\end{split}
\end{equation}
where the integrand is also summarized from (\ref{eq4-17}).

Furthermore, it can be readily shown that (\ref{eq4-18}) is also true for $\left( {Q + 1} \right)$ training periods. More specifically, we have
\begin{equation}\label{eq4-19}
    \begin{split}
&\int\limits_0^\pi  {\int\limits_0^{{\theta _{Q + 1}}} {\left\{ {\sum\limits_{i = 1}^{\left\lceil {{{\left( {Q - 2} \right)} \mathord{\left/
 {\vphantom {{\left( {Q - 2} \right)} 2}} \right.
 \kern-\nulldelimiterspace} 2}} \right\rceil } {\frac{{{{\left( { - 1} \right)}^{i + 1}}\theta _Q^{Q - 1 - 2i}}}{{\left( {Q - 1 - 2i} \right)!}}}  + \cos \left( {{\theta _Q} - \frac{{\left( {Q - 1} \right)\pi }}{2}} \right)} \right\}d{\theta _Q}} } d{\theta _{Q + 1}}\\
& = \int\limits_0^\pi  {\left\{ {\sum\limits_{i = 1}^{\left\lceil {{{\left( {Q - 2} \right)} \mathord{\left/
 {\vphantom {{\left( {Q - 2} \right)} 2}} \right.
 \kern-\nulldelimiterspace} 2}} \right\rceil } {\frac{{{{\left( { - 1} \right)}^{i + 1}}\theta _{Q + 1}^{Q - 2i}}}{{\left( {Q - 2i} \right)!}}}  + \sin \left( {{\theta _{Q + 1}} - \frac{{\left( {Q - 1} \right)\pi }}{2}} \right) + \sin \left( {\frac{{\left( {Q - 1} \right)\pi }}{2}} \right)} \right\}} d{\theta _{Q + 1}}\\
 &= \sum\limits_{i = 1}^{\left\lceil {{Q \mathord{\left/
 {\vphantom {Q 2}} \right.
 \kern-\nulldelimiterspace} 2}} \right\rceil } {{{\left( { - 1} \right)}^{i + 1}}f\left( {Q + 1 - 2i} \right){\pi ^{Q + 1 - 2i}}}.
\end{split}
\end{equation}

As a result, the general formula of $\mathbb{E}\left\{ {\mathop {\max }\limits_{q = 1,2, \cdots ,Q} \left\{ {\cos {\theta _q}} \right\}} \right\}$ can be simplified to
\begin{equation}\label{eq4-20}
    \mathbb{E}\left\{ {\mathop {\max }\limits_{q = 1,2, \cdots ,Q} \left\{ {\cos {\theta _q}} \right\}} \right\} = \frac{{Q!}}{{{\pi ^Q}}}\sum\limits_{i = 1}^{\left\lceil {{{\left( {Q - 1} \right)} \mathord{\left/
 {\vphantom {{\left( {Q - 1} \right)} 2}} \right.
 \kern-\nulldelimiterspace} 2}} \right\rceil } {{{\left( { - 1} \right)}^{i + 1}}f\left( {Q - 2i} \right){\pi ^{Q - 2i}}}.
\end{equation}

On the other hand, considering that both ${h_r}$ and ${h_d}$ obey complex Gaussian distributions, we can show that
\begin{equation}\label{eq4-21}
\mathbb{E}\left\{ {{{\left| {{h_r}} \right|}^2}} \right\} = \rho _r^2,\quad \mathbb{E}\left\{ {{{\left| {{h_d}} \right|}^2}} \right\} = \rho _d^2,\mathbb{E}\left\{ {\left| {{h_r}} \right|} \right\} = \frac{{\sqrt \pi  }}{2}{\rho _r},\quad \mathbb{E}\left\{ {\left| {{h_d}} \right|} \right\} = \frac{{\sqrt \pi  }}{2}{\rho _d}.
\end{equation}

Upon substituting (\ref{eq4-20}) and (\ref{eq4-21}) into (\ref{eq4-13}), the proof is completed. $\hfill\blacksquare$

\section{Proof of \textbf{\emph{Proposition 2}}}\label{A2}
Firstly, the channel gain in (\ref{eq4-22}) is upper bounded by
\begin{equation}\label{eq4-24}
{\mathbb{E}}\left\{ {\mathop {\max }\limits_{q = 1,2, \cdots ,Q} \left\{ {{{\left| {\sum\limits_{n = 1}^N {{\varphi _{q,n}}h_{r,n}^H}  + h_d^H} \right|}^2}} \right\}} \right\} \le {\mathbb{E}}\left\{ {{{\left| {\sum\limits_{n = 1}^N {{{\hat \varphi }_{q,n}}h_{r,n}^H}  + h_d^H} \right|}^2}} \right\},
\end{equation}
where ${\hat \varphi _{q,n}}$ represents the optimal configuration of ${\varphi _{q,n}}$ for $Q$ training periods, when only the $n$th reflecting element is turned on, while all others are muted, which can be expressed as
\begin{equation}\label{eq4-25}
{\hat \varphi _{q,n}} = \arg \mathop {\max }\limits_{{\varphi _{q,n}},q = 1,2, \cdots ,Q} \left\{ {{{\left| {{\varphi _{q,n}}h_{r,n}^H + h_d^H} \right|}^2}} \right\}.
\end{equation}

Furthermore, the right-hand-side (RHS) of (\ref{eq4-25}) can be expanded as
\begin{equation}\label{eq4-26}
    \begin{split}
{\mathbb{E}}\left\{ {{{\left| {\sum\limits_{n = 1}^N {{{\hat \varphi }_{q,n}}h_{r,n}^H}  + h_d^H} \right|}^2}} \right\}&= \sum\limits_{n = 1}^N {{\mathbb{E}}\left\{ {{{\left| {h_{r,n}^H} \right|}^2}} \right\}}  + {\mathbb{E}}\left\{ {{{\left| {h_d^H} \right|}^2}} \right\}+ \frac{\pi }{2}N\sum\limits_{n = 1}^N {{\mathbb{E}}\left\{ {\left| {h_{r,n}^H} \right|} \right\}{\mathbb{E}}\left\{ {\left| {h_d^H} \right|} \right\}{\mathbb{E}}\left\{ {\cos {{\hat \theta }_n}} \right\}} \\
 &+ \frac{\pi }{4}\sum\limits_{n = 1}^N {\sum\limits_{n' = 1,n' \ne n}^N {{\mathbb{E}}\left\{ {\left| {h_{r,n}^H} \right|} \right\}{\mathbb{E}}\left\{ {\left| {h_{r,n'}^H} \right|} \right\}{\mathbb{E}}\left\{ {\cos \left( {{{\hat \theta }_n} - {{\hat \theta }_{n'}}} \right)} \right\}} },
\end{split}
\end{equation}
where ${\hat \theta _n} = \arg \mathop {\max }\limits_{{\theta _{q,n}},q = 1,2, \cdots ,Q} \left\{ {\cos {\theta _{q,n}}} \right\}$ is the one associated with the largest cosine of ${\theta _{q,n}} = \arccos \frac{{\Re \left( {\varphi _{q,n}^ * {h_{r,n}}h_d^H} \right)}}{{\left| {{h_{r,n}}} \right| \cdot \left| {h_d^H} \right|}},q = 1,2, \cdots ,Q$.

Following the same philosophy as in (\ref{eq4-20}), we have
\begin{equation}\label{eq4-27-0}
    {\mathbb{E}}\left\{ {\cos {{\hat \theta }_n}} \right\}=g\left( Q \right),\quad n = 1,2, \cdots ,N,
\end{equation}
while ${\mathbb{E}}\left\{ {\cos \left( {{{\hat \theta }_n} - {{\hat \theta }_{n'}}} \right)} \right\}$ can be expressed as
\begin{equation}\label{eq4-27}
    \begin{split}
{\mathbb{E}}\left\{ {\cos \left( {{{\hat \theta }_n} - {{\hat \theta }_{n'}}} \right)} \right\}&= {\mathbb{E}}\left\{ {\cos {{\hat \theta }_n}\cos {{\hat \theta }_{n'}} + \sin {{\hat \theta }_n}\sin {{\hat \theta }_{n'}}} \right\}\\
 &= {\mathbb{E}}\left\{ {\mathop {\max }\limits_{q = 1,2, \cdots ,Q} \left\{ {\cos {{\theta }_{q,n}}} \right\}} \right\}{\mathbb{E}}\left\{ {\mathop {\max }\limits_{q = 1,2, \cdots ,Q} \left\{ {\cos {{\theta }_{q,n'}}} \right\}} \right\}= {g^2}\left( Q \right),
\end{split}
\end{equation}
where we have ${\mathbb{E}}\left\{ {\sin {{\hat \theta }_n}} \right\}=0$ for $n = 1,2, \cdots ,N$ due to the anti-symmetric nature of the sine function with respect to the real axis.

Upon substituting (\ref{eq4-27-0}) and (\ref{eq4-27}) into (\ref{eq4-26}), we complete the proof. $\hfill\blacksquare$

\section{Proof of \textbf{\emph{Proposition 3}}}\label{A3}

When the equi-partition configuration scheme of Fig. \ref{f4} is adopted, $\mathbb{E}\left\{ {\mathop {\max }\limits_{q = 1,2, \cdots ,Q} \left\{ {\cos {\theta _q}} \right\}} \right\}$ of (\ref{eq4-13}) can be calculated as
\begin{equation}\label{eq4-34}
\mathbb{E}\left\{ {\mathop {\max }\limits_{q = 1,2, \cdots ,Q} \left\{ {\cos {\theta _q}} \right\}} \right\} = \frac{Q}{{2\pi }}\int\limits_{ - {\pi  \mathord{\left/
 {\vphantom {\pi  Q}} \right.
 \kern-\nulldelimiterspace} Q}}^{{\pi  \mathord{\left/
 {\vphantom {\pi  Q}} \right.
 \kern-\nulldelimiterspace} Q}} {\cos {\theta _1}d{\theta _1}}  = \frac{{\sin \left( {{\pi  \mathord{\left/
 {\vphantom {\pi  Q}} \right.
 \kern-\nulldelimiterspace} Q}} \right)}}{{\left( {{\pi  \mathord{\left/
 {\vphantom {\pi  Q}} \right.
 \kern-\nulldelimiterspace} Q}} \right)}}.
\end{equation}

Upon substituting (\ref{eq4-34}) into (\ref{eq4-26}) and following the same considerations as for (\ref{eq4-27-0}) and (\ref{eq4-27}), we complete the proof. $\hfill\blacksquare$

\section{Proof of \textbf{\emph{Proposition 4}}}\label{A4}
According to (\ref{eq5-2}), the average channel gain of (\ref{eq4-26}) in the presence of channel estimation errors can be expressed as
\begin{equation}\label{eq5-4}
    \begin{split}
&\mathbb{E}\left\{ {{{\left| {\sum\limits_{n = 1}^N {{{\hat \varphi }_n}{h_{r,n}}}  + h_d} \right|}^2}} \right\}\\
&= \mathbb{E}\left\{ {{{\left| {\sum\limits_{n = 1}^N {\frac{{\hat h_{r,n}^*{{\hat h}_d}}}{{\left| {\hat h_{r,n}^*{{\hat h}_d}} \right|}}{h_{r,n}}}  + h_d} \right|}^2}} \right\}\\
 &= \mathbb{E}\left\{ {{{\left| {\sum\limits_{n = 1}^N {\frac{{\left( {h_{r,n}^* + \varepsilon _{r,n}^*} \right)}}{{\left| {\left( {h_{r,n}^* + \varepsilon _{r,n}^*} \right)} \right|}}{h_{r,n}}}  + \frac{{\left( {h_d^* + \varepsilon _d^*} \right)}}{{\left| {\left( {h_d^* + \varepsilon _d^*} \right)} \right|}}{h_d}} \right|}^2}} \right\}\\
 &= \mathbb{E}\left\{ {\sum\limits_{n = 1}^N {{{\left| {{h_{r,n}}} \right|}^2}} } \right\} + \mathbb{E}\left\{ {{{\left| {{h_d}} \right|}^2}} \right\}\\
 &+ \mathbb{E}\left\{ {2\sum\limits_{n = 1}^N {\Re \left( {{{\left( {\frac{{\left( {h_{r,n}^* + \varepsilon _{r,n}^*} \right)}}{{\left| {\left( {h_{r,n}^* + \varepsilon _{r,n}^*} \right)} \right|}}{h_{r,n}}} \right)}^*}\frac{{\left( {h_d^* + \varepsilon _d^*} \right)}}{{\left| {\left( {h_d^* + \varepsilon _d^*} \right)} \right|}}{h_d}} \right)} } \right\}\\
 &+ \mathbb{E}\left\{ {{\rm{2}}\sum\limits_{n = 1}^{N - 1} {\sum\limits_{n' = n + 1}^N {\Re \left( {{{\left( {\frac{{\left( {h_{r,n}^* + \varepsilon _{r,n}^*} \right)}}{{\left| {\left( {h_{r,n}^* + \varepsilon _{r,n}^*} \right)} \right|}}{h_{r,n}}} \right)}^*}\left( {\frac{{\left( {h_{r,n'}^* + \varepsilon _{r,n'}^*} \right)}}{{\left| {\left( {h_{r,n'}^* + \varepsilon _{r,n'}^*} \right)} \right|}}{h_{r,n'}}} \right)} \right)} } } \right\}.
\end{split}
\end{equation}

Furthermore, the $n$th entry of the second part of the RHS of (\ref{eq5-4}) can be expanded as
\begin{equation}\label{eq5-5}
    \begin{split}
&\mathbb{E}\left\{ {\Re \left( {{{\left( {\frac{{\left( {h_{r,n}^* + \varepsilon _{r,n}^*} \right)}}{{\left| {\left( {h_{r,n}^* + \varepsilon _{r,n}^*} \right)} \right|}}{h_{r,n}}} \right)}^*}\frac{{\left( {h_d^* + \varepsilon _d^*} \right)}}{{\left| {\left( {h_d^* + \varepsilon _d^*} \right)} \right|}}{h_d}} \right)} \right\}\\
 &= \mathbb{E}\left\{ {\Re \left( {{{\left( {\frac{{\left( {h_{r,n}^* + \varepsilon _{r,n}^*} \right)}}{{\left| {\left( {h_{r,n}^* + \varepsilon _{r,n}^*} \right)} \right|}}{h_{r,k}}} \right)}^*}} \right)\Re \left( {\frac{{\left( {h_d^* + \varepsilon _d^*} \right)}}{{\left| {\left( {h_d^* + \varepsilon _d^*} \right)} \right|}}{h_d}} \right)} \right\}\\
 &- \mathbb{E}\left\{ {\Im \left( {{{\left( {\frac{{\left( {h_{r,n}^* + \varepsilon _{r,n}^*} \right)}}{{\left| {\left( {h_{r,n}^* + \varepsilon _{r,n}^*} \right)} \right|}}{h_{r,k}}} \right)}^*}} \right)\Im \left( {\frac{{\left( {h_d^* + \varepsilon _d^*} \right)}}{{\left| {\left( {h_d^* + \varepsilon _d^*} \right)} \right|}}{h_d}} \right)} \right\}.
\end{split}
\end{equation}

Owing to the assumptions of ${h_{r,n}} \sim {\cal C}{\cal N}\left( {0,\rho _r^2} \right),n = 1,2, \cdots ,N$, ${h_d} \sim \mathcal{CN}\left( {0,\rho _d^2} \right)$, ${\varepsilon _d} \sim {\mathcal{CN}}\left( {0,\sigma _d^2} \right)$ and ${\varepsilon _{r,n}} \sim {\mathcal{CN}}\left( {0,\sigma _r^2} \right),n = 1,2, \cdots ,N$, we have
\begin{equation}\label{eq5-6}
\mathbb{E}\left\{ {\Re \left( {{{\left( {\frac{{\left( {h_{r,n}^* + \varepsilon _{r,n}^*} \right)}}{{\left| {\left( {h_{r,n}^* + \varepsilon _{r,n}^*} \right)} \right|}}{h_{r,k}}} \right)}^*}} \right)} \right\} = \frac{{\sqrt \pi  \rho _{r,n}^2}}{{2\sqrt {\rho _{r,n}^2 + \sigma _{r,n}^2} }},
\end{equation}
\begin{equation}\label{eq5-7}
\mathbb{E}\left\{ {\Re \left( {{{\left( {\frac{{\left( {h_d^* + \varepsilon _d^*} \right)}}{{\left| {\left( {h_d^* + \varepsilon _d^*} \right)} \right|}}{h_d}} \right)}^*}} \right)} \right\} = \frac{{\sqrt \pi  \rho _d^2}}{{2\sqrt {\rho _d^2 + \sigma _d^2} }},
\end{equation}
\begin{equation}\label{eq5-8}
\mathbb{E}\left\{ {\Im \left( {{{\left( {\frac{{\left( {h_{r,n}^* + \varepsilon _{r,n}^*} \right)}}{{\left| {\left( {h_{r,n}^* + \varepsilon _{r,n}^*} \right)} \right|}}{h_{r,k}}} \right)}^*}} \right)} \right\} = 0,
\end{equation}
\begin{equation}\label{eq5-9}
\mathbb{E}\left\{ {\Im \left( {{{\left( {\frac{{\left( {h_d^* + \varepsilon _d^*} \right)}}{{\left| {\left( {h_d^* + \varepsilon _d^*} \right)} \right|}}{h_d}} \right)}^*}} \right)} \right\} = 0.
\end{equation}

Following the derivatives of (\ref{eq5-6}) $\sim$ (\ref{eq5-9}) for the last term in (\ref{eq5-5}), we arrive at:
\begin{equation}\label{eq5-10}
{\mathbb{E}}\left( {{{\left| {\sum\limits_{n = 1}^N {{{\hat \varphi }_n}{h_{r,n}}}  + {h_d}} \right|}^2}} \right) = N\rho _r^2 + \rho _d^2 + \frac{{\pi N\rho _r^2\rho _d^2}}{{2\sqrt {\left( {\rho _r^2 + \sigma _r^2} \right)\left( {\rho _d^2 + \sigma _d^2} \right)} }} + \frac{{\pi N\left( {N - 1} \right)\rho _r^4}}{{4\left( {\rho _r^2 + \sigma _r^2} \right)}}.
\end{equation}

Substituting (\ref{eq5-10}) into (\ref{eq4-22}), we complete the proof. $\hfill\blacksquare$

\bibliography{An}
\end{document}